\documentclass[fleqn,usenatbib]{mnras}
\usepackage{newtxtext,newtxmath}
\usepackage[T1]{fontenc}
\usepackage{mwe}

\DeclareRobustCommand{\VAN}[3]{#2}
\let\VANthebibliography\thebibliography
\def\thebibliography{\DeclareRobustCommand{\VAN}[3]{##3}\VANthebibliography}

\usepackage{amsmath}	% Advanced maths commands
\usepackage{graphicx}
\usepackage{tikz}
\usepackage{gensymb}
\usepackage{glossaries}
\usepackage{lineno}
\usepackage{subcaption}
\usepackage{pdflscape}

\setacronymstyle{long-short}
\glsdisablehyper

\graphicspath{{./},{./figures/}}

\usepackage{csvsimple}

\newcommand{\annotate}[2]{\begin{tikzpicture}
    \node[anchor=south west,inner sep=0,align=center] (image) at (0,0) {
    #1
    };
    \begin{scope}[x={(image.south east)},y={(image.north west)}]
    #2
    \end{scope}
\end{tikzpicture}}

\newcommand{\numax}{\nu_{\mathrm{max}}}

\newcommand{\dnu}{\delta\nu}
\newcommand{\Dnu}{\Delta\nu}
\newcommand{\DPi}{\Delta\Pi}

\newcommand{\dnurot}{\dnu_{\mathrm{rot}}}
\newcommand{\dnumag}{\dnu_{\mathrm{mag}}}

\newacronym{PSD}{PSD}{power spectral density}

\newacronym{RG}{RG}{red giant}

\newacronym{S/N}{S/N}{signal-to-noise ratio}

\newcommand{\Teff}{\,T_\mathrm{eff}}
\newcommand{\Msun}{\,\mathrm{M}_\odot}

\newcommand{\Brms}{\langle \mathrm{B}_{\mathrm{r}}^2 \rangle}

\usepackage{soul}

\title[Rotation and magnetic perturbations on the RGB]{Asteroseismic Signatures of Core Magnetism and Rotation in Hundreds of Low-Luminosity Red Giants.}

\author[Hatt E. J. et al]{
Emily J. Hatt,$^{1}$\thanks{E-mail: exh698@student.bham.ac.uk},
J. M. Joel Ong $^{2}$\thanks{Hubble Fellow},
Martin B. Nielsen $^{1}$,
William J. Chaplin $^{1}$,
Guy R. Davies $^{1}$,\newauthor 
S\'ebastien Deheuvels $^{3}$,
Jérôme Ballot$^{3}$,
Gang Li$^{3,4}$,
Lisa Bugnet $^{5}$\\
$^{1}$Royal Astronomical Society, Burlington House, Piccadilly, London W1J 0BQ, UK\\
$^{2}$ Institute for Astronomy, University of Hawai‘i, 2680 Woodlawn Drive, Honolulu, HI 96822, USA\\
$^{3}$ IRAP, Universit\'e de Toulouse, CNRS, CNES, UPS, 14 avenue Edouard Belin, 31400 Toulouse, France\\
$^{4}$ Institute of Astronomy, KU Leuven, Celestijnenlaan 200D, 3001 Leuven, Belgium\\
${^5}$ Institute of Science and Technology Austria (IST Austria), Am Campus 1, Klosterneuburg, Austria \\}

\date{Accepted XXX. Received YYY; in original form ZZZ}

\pubyear{2024}

\begin{document}
\label{firstpage}
\pagerange{\pageref{firstpage}--\pageref{lastpage}}
\maketitle

% Abstract of the paper
\begin{abstract}
Red Giant stars host solar-like oscillations which have mixed character, being sensitive to conditions both in the outer convection zone and deep within the interior. 
The properties of these modes are sensitive to both core rotation and magnetic fields. While asteroseismic studies of the former have been done on a large scale, studies of the latter are currently limited to tens of stars. We aim to produce the first large catalogue of both magnetic and rotational perturbations. We jointly constrain these parameters by devising an automated method for fitting the power spectra directly. We successfully apply the method to 302 low-luminosity red giants. We find a clear bimodality in core rotation rate. The primary peak is at $\dnurot$ = 0.32 $\mu$Hz, and the secondary at $\dnurot$ = 0.47 $\mu$Hz. Combining our results with literature values, we find that the percentage of stars rotating much more rapidly than the population average increases with evolutionary state. We measure magnetic splittings of 2$\sigma$ significance  in 23 stars. While the most extreme magnetic splitting values appear in stars with masses > 1.1M$_{\odot}$, implying they formerly hosted a convective core, a small but statistically significant magnetic splitting is measured at lower masses. Asymmetry between the frequencies of a rotationally split multiplet has previously been used to diagnose the presence of a magnetic perturbation. We find that of the stars with a significant detection of magnetic perturbation, 43\% do not show strong asymmetry. We find no strong evidence of correlation between the rotation and magnetic parameters.
\end{abstract}

% Select between one and six entries from the list of approved keywords.
% Don't make up new ones.

%%%%%%%%%%%%%%%%%%%%%%%%%%%%%%%%%%%%%%%%%%%%%%%%%%

%%%%%%%%%%%%%%%%% BODY OF PAPER %%%%%%%%%%%%%%%%%%

\section{Introduction}\label{introduction}

Magnetic fields play a critical role in stellar evolution. The prevailing theory regarding those which are observed in the Sun and other main-sequence solar-type stars is that they are generated by a dynamo process. This mechanism is, crucially, dependent on the interplay between turbulent convection and differential rotation \citep{1984ApJ...279..763N}. Convection is supported at various phases in the lifecycle of low to intermediate mass stars, providing several avenues for the formation of a magnetic field. Magnetic fields have been invoked as a possible solution to many open problems in stellar evolution (see \citealt{2017LRSP...14....4B} for a review). Notably, a magnetic field could transport angular momentum from the core to the outer envelope, reducing the degree of differential rotation occurring in evolved stars \citep{2014ApJ...788...93C, 2016A&A...589A..23S, 2019A&A...631L...6E, 2019MNRAS.485.3661F, 2022A&A...661A.119G, 2022A&A...664L..16E,2023A&A...673A.110M}. This could resolve the observed discrepancy between predicted rotation rates in the cores of evolved stars and those observed, the latter being (at best) two orders of magnitude too small \citep{2012A&A...544L...4E, 2013A&A...549A..74M, 2013A&A...555A..54C}.

Given magnetic fields in the outer layers of stars can be detected in light curves due to the manifestation of this field as star-spots, most detected stellar magnetic fields in convection zones are sun-like in nature. However, a magnetic field capable of producing observed red giant branch (RGB) rotation rates would need to operate near the stellar core \citep{2014ApJ...793..123M}. Core convection is expected in stars with masses above $\approx$1.1M$_{\odot}$ during the main sequence \citep{1990sse..book.....K}. It is possible for these fields to remain stable as the star evolves off of the main sequence, where the interior becomes radiative \citep{2017ApJ...846....8E,2019A&A...622A..72V,2022MNRAS.511..732B}. Even without the presence of a convective core on the main sequence, it is possible to create a magnetic field in stably-stratified zones via a Tayler-Spruit dynamo or related processes \citep{2002A&A...381..923S, 2019MNRAS.485.3661F, 2022A&A...664L..16E, 2023Sci...379..300P}. 

Asteroseismology, the study of stellar pulsations, offers the only probe sensitive to near core regions. Solar-like oscillations come in two types, propagating in two largely distinct regions. In the surface convection zones, turbulent motion drives acoustic oscillations known as pressure or p-modes. Closer to the core, strong density stratification supports bouyancy oscillations, gravity or g-modes. When the frequencies of p and g-modes approach each other, the two types of modes can couple forming what is known as a mixed-mode \citep{1989nos..book.....U}. Sharing properties of both the pure p and g modes, mixed-modes are both sensitive to conditions near the stellar core and the surface. Mode amplitudes reach a maximum about a characteristic oscillation frequency ($\nu_{\mathrm{max}}$) that scales with the acoustic cut off. On the main sequence, the maximum g-mode frequency is significantly lower than $\nu_{\mathrm{max}}$, such that mixed modes are not excited to observable amplitudes. As a star evolves off of the main sequence onto the RGB, $\nu_{\mathrm{max}}$ decreases in response to the expansion of the outer layers. Concurrently the core contracts, increasing the density of g-modes. As $\nu_{\mathrm{max}}$ approaches the maximum frequency of the g-modes, mixed modes become increasingly observable. The $\textit{Kepler}$ telescope \citep{2010Sci...327..977B} observed a large number of evolved stars with high-precision photometry, such that we are able to measure mixed modes in thousands of stars \citep{2014A&A...572L...5M, 2016A&A...588A..87V,2023ApJ...954..152K}.

The presence of a magnetic field within the region where modes propagate has been shown to perturb mode frequencies \citep{1990MNRAS.242...25G, 1992ApJ...395..307G, 1994PASJ...46..301T, 2005A&A...444L..29H, 2021A&A...647A.122M, 2021A&A...650A..53B, li_magnetic_2022, 2023A&A...676L...9M}. This owes both directly to the introduction of the Lorentz force into the equations of stellar oscillation and indirectly by impacting the properties of mode cavities. Unlike rotation, magnetic fields are not, in general, azimuthally symmetric. As such, the degree to which a mode is perturbed is dependent both on field strength and its geometry and topology \citep{2020MNRAS.496..620G, 2021A&A...647A.122M, 2021A&A...650A..53B, 2021MNRAS.504.3711L}. Alongside identifying the presence of a magnetic field, measurements of perturbations to mode frequencies can put key constraints on field strength and structure. The theoretical tools required to exploit the perturbed spectra of evolved stars in such a way have only just been established. Furthermore, the size of the parameter space involved with fitting even unperturbed mixed modes makes the problem computationally expensive and contingent on well-informed priors \citep{2023ApJ...954..152K}. Even more free parameters are required when considering perturbations, amplifying the issue. As such, only the cases with the strongest magnetic signatures have thus far been analysed. Accordingly current catalogues of core magnetic fields identified through perturbations to mode frequencies are very limited, numbering 24 stars at time of writing \citep{li_magnetic_2022, 2023A&A...680A..26L, 2023A&A...670L..16D}. 

In addition to perturbing frequencies, it has been shown that once the field strength exceeds a critical value (typically in the range of 10$^5$ - 10$^7$G for a low-luminosity red giant) it can act to suppress the amplitudes of mixed dipole modes \citep{2015Sci...350..423F}. A number of red giants observed by \textit{Kepler} exhibit mode amplitudes that suggest they have been altered by such a field \citep{2014A&A...563A..84G, 2015Sci...350..423F}. The identification of these depressed dipole modes implies a field strength in the core which must exceed the critical value. In this work we investigate perturbations to the mode frequencies, such that we do not select stars in which the dipole modes are substantially depressed. Accordingly we are sensitive to intermediate field strengths, likely below the critical value defined in \citet{2015Sci...350..423F}.

We investigate the perturbations caused by core rotation and magnetic fields using a sample of 302 low luminosity RGB stars observed by \textit{Kepler}. In section \ref{sec: methods} we describe how the sample is selected from the > 16000 RGB stars in \cite{yu_asteroseismology_2018} (hereafter Y18). We then go on to fit a perturbed asymptotic expression to the power spectra. To enable a large scale fitting without the problem becoming computationally intractable, we construct priors on the perturbed quantities. This is done via a novel method of exploiting stretched period \'echelle diagrams calculated using the spectrum directly, a tool so far only used on previously measured mode frequencies. In section \ref{sec:results} we detail the resulting measurements, before discussing correlations with fundamental stellar properties in section \ref{sec:discuss}.

\section{Survey Methodology}\label{sec: methods}

\subsection{Target Selection}\label{Target Selection}
\begin{figure}
    \centering
    \includegraphics[width=1\linewidth]{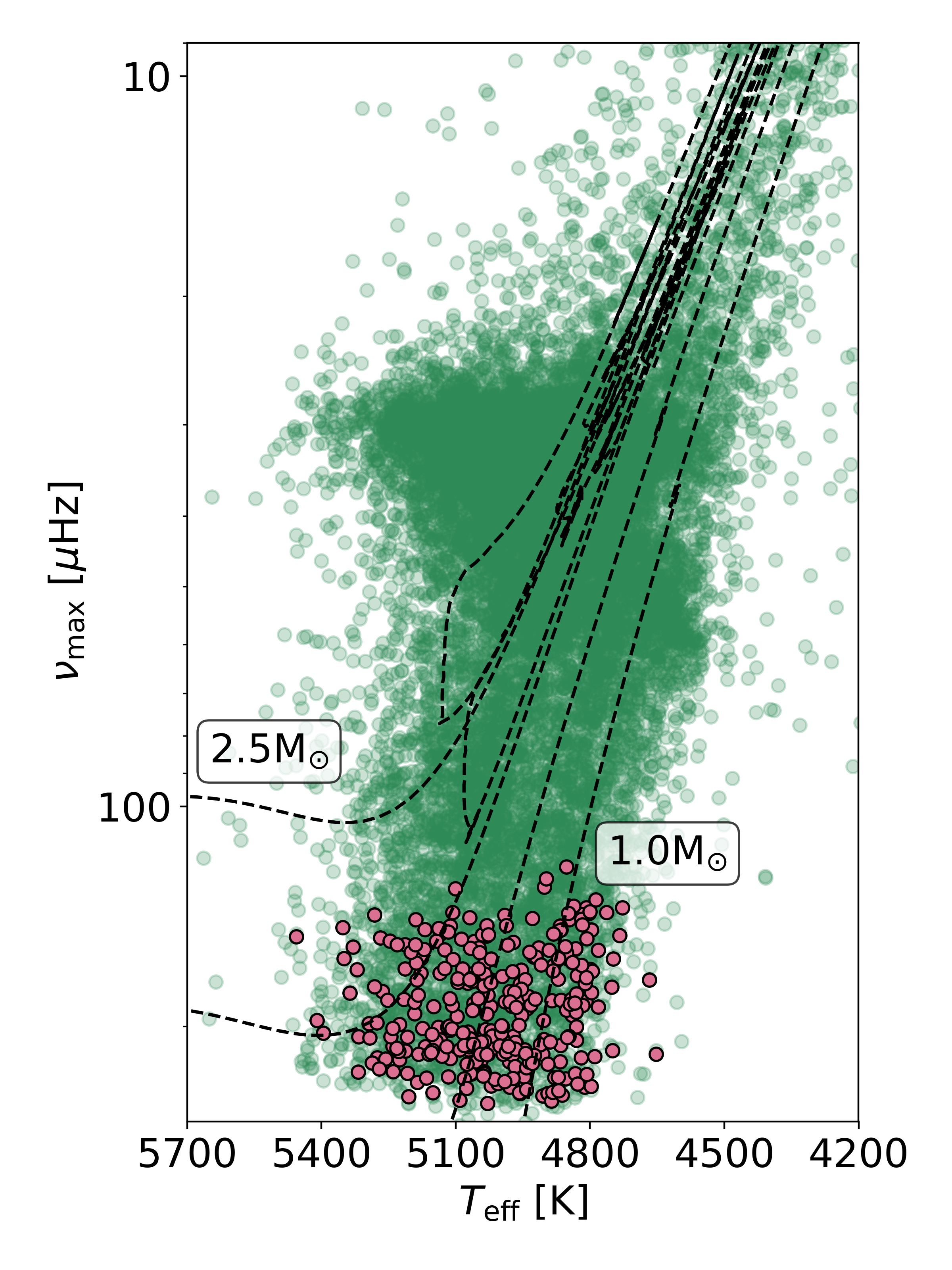}
    \caption{$\numax$ and $\Teff$ values for the selected sample (pink) compared to those from the catalogue by \citet{yu_asteroseismology_2018} (green). For reference we show the MIST \citep{2016ApJ...823..102C, 2016ApJS..222....8D} evolutionary tracks for $1\Msun$, $1.5\Msun$, $2\Msun$, and $2.5\Msun$ stars at $[\mathrm{Fe/H}]=-0.25$ which approximately corresponds to the median metallicity of the selected sample, as reported in table 2 of \citet{yu_asteroseismology_2018}.}
    \label{fig:targets}
\end{figure}
We used $\textit{Kepler}$ long-cadence light curves, calculating the power spectral density via a Lomb-Scargle periodogram \citep{1976Ap&SS..39..447L, 1982ApJ...263..835S} using the \texttt{lightkurve} package \citep{2018ascl.soft12013L}.

Of the observable mixed modes present in a given spectrum, those which are gravity-dominated are the most sensitive to core conditions. As the widths of these modes are smaller than their pressure-dominated counterparts \citep{2011A&A...525L...9M, mosser_period_2015, mosser_period_2018, 2016A&A...588A..87V}, they set an upper limit on the frequency resolution we required. For such modes, the linewidths are on the order of 0.01$\mu$Hz \citep{yu_asteroseismology_2018, 2020MNRAS.495.2363L} necessitating time series that exceed three years in length. Therefore we restrict ourselves to stars that were observed by \textit{Kepler} for a full 4-years.

For a given radial order and $\ell$, rotation breaks the degeneracy between modes of differing azimuthal order, $m$. To leading order in perturbation theory (i.e. for slow rotation), the degree to which a mode is perturbed by rotation is proportional to $m$, such that modes of order $m$ = 0 are not affected, remaining at the unperturbed frequency. For modes with non-zero $m$, the magnitude of the perturbation is shared between two modes having azimuthal orders $m$ and $m'$ if $|m|$ = $|m'|$. The sign of $m$ then determines whether the mode increases or decreases in frequency. 

Magnetic fields will perturb all components of a multiplet, with (generally) a different shift for all $|m|$ components. The resulting asymmetry in the spectrum of a magnetically perturbed multiplet is distinct, and as such is commonly used to establish a detection. However, the degree to which the shift differs among components is dependent on the topology of the field (see section \ref{sec: dipole model}) and certain configurations result in zero asymmetry regardless of field strength \citep{2021MNRAS.504.3711L,2023A&A...676L...9M, li_magnetic_2022}. Therefore, we make no selection based on asymmetry but rather focus on the average shift of the multiplet. In the following we used modes of angular degree $\ell$ = 1 (dipole modes, see section \ref{sec: dipole model}), such that multiplets consist of 3 components ($m = 0, \pm 1$). Given all components of the multiplet contain information about the magnetism, we restricted ourselves to targets where-in we could visually identify all three peaks for a given $\ell$ = 1 mode. This occurs at intermediate inclination, approximately in the range 30$\degree$ < $i$ < 60$\degree$.

Finally, as a star evolves along the RGB, the mixed mode density ($\mathcal{N}$ = $\Dnu/\DPi_{1} \numax^2$) increases. Once the rotational splitting is of the order of the separation between adjacent mixed modes, identification of multiplet components becomes difficult. Therefore we restrict ourselves to targets with $\numax$ > 100 $\mu$Hz  (implying mixed mode densities in the region of $\mathcal{N}$ < 10). The upper limit on $\numax$ is set by the Nyquist limit in the data (277 $\mu$Hz). Applying these constraints to the > 16000 targets identified in Y18, we construct a list of 334 stars (see Fig. \ref{fig:targets}). 

\subsection{Data pre-processing} 
Rather than measuring mode frequencies prior to fitting for the perturbed quantities, we fit the power spectra directly. Given p-modes of angular degree $\ell$ = 0 do not couple to g-modes, they will not provide information about core magnetism or rotation. Additionally, mode coupling in modes of angular degree $\ell$ = 2 is orders of magnitude smaller than that in $\ell$ = 1 modes. As such, we removed the additional power from the $\ell$ = 0 and 2 modes (see sections \ref{sec: S/N}, \ref{sec: 2,0}) prior to fitting a perturbed expression to the spectrum of the dipole modes (section \ref{sec: dipole model}).

\subsubsection{Computing the S/N spectrum} \label{sec: S/N}
The first step in the process is to estimate the background noise level. Here we define the background noise as any power that is not directly attributed to the oscillation modes. For stars on the red giant branch (RGB) the background typically consists of a frequency independent term due to photon noise, two frequency dependent terms due to granulation on the stellar surface, and finally a third frequency dependent term which accounts for any residual, long-term, instrumental variability \citep[see, e.g,][]{Kallinger2014}. 

Here we model the photon noise as a frequency independent, random, variable. Three frequency-dependent terms as Harvey-like profiles \citep{Harvey1985}, following \citet{Kallinger2014}, are introduced to model the signature of granulation. All the terms in the background noise model vary slowly with frequency, and so we bin the spectrum in log-frequency, after which the model parameters are sampled. 

We evaluate a set of $100$ draws from the model posterior distribution to compute a mean background level on the unbinned frequency grid. We divide the power spectral density (PSD) by the mean background model to obtain a residual S/N spectrum which now only contains the oscillation envelope.

\subsubsection{Establishing a mean $\ell=2,0$ model}\label{sec: 2,0}

The next step is to remove the contribution of the $\ell=0$ and $\ell=2$ modes to the S/N spectrum. This is done by computing a mean $\ell=2,0$ model which is then used to obtain a residual S/N spectrum which notionally only contains the $\ell=1$ modes. The $\ell=2,0$ model that we use is consistent with that of \citet{Nielsen2021} which, to summarize, consists of a set of mode frequencies determined by the asymptotic p-mode relation for the $\ell=0$ and $\ell=2$ modes. The spectrum is then approximated as a sum of Lorentzian profiles at these frequencies, modulated by a Gaussian envelope in power setting the mode heights. For simplicity, we set a single width for all modes. To construct the mean $\ell=2,0$ model we draw 50 samples from the model posterior distribution, and average the resulting model spectra. The sampling is performed using a principal component based dimensionality reduction method presented in \citet{Nielsen2023}. 

Dividing the S/N spectrum by the mean $\ell=2,0$ model leaves us with a residual spectrum which consists primarily of power due to the $\ell=1$ modes, any potential $\ell=3$ modes, and to a lesser extent any residual power remaining due to errors in the $\ell=2,0$ model. While this simplifies the sampling of the $\ell=1$ model posterior distribution, the inference on the core rotation and magnetism is now conditional on the background and $\ell=2,0$ models. This means that in the following we neglect any errors due to uncertainty in these models. However, for high S/N red giant stars where the background and $\ell=2,0$ model parameters can be precisely estimated, this is not expected to be a significant contribution to uncertainty on the rotation and magnetic field terms. We also cannot capture correlations between perturbed quantities and the background. 

\subsection{Estimating the $\ell=1$ model parameters}\label{sec: dipole model}

\begin{figure}
    \centering
    \annotate{\includegraphics[trim=0 1.2cm 0 0, clip, width=.45\textwidth]{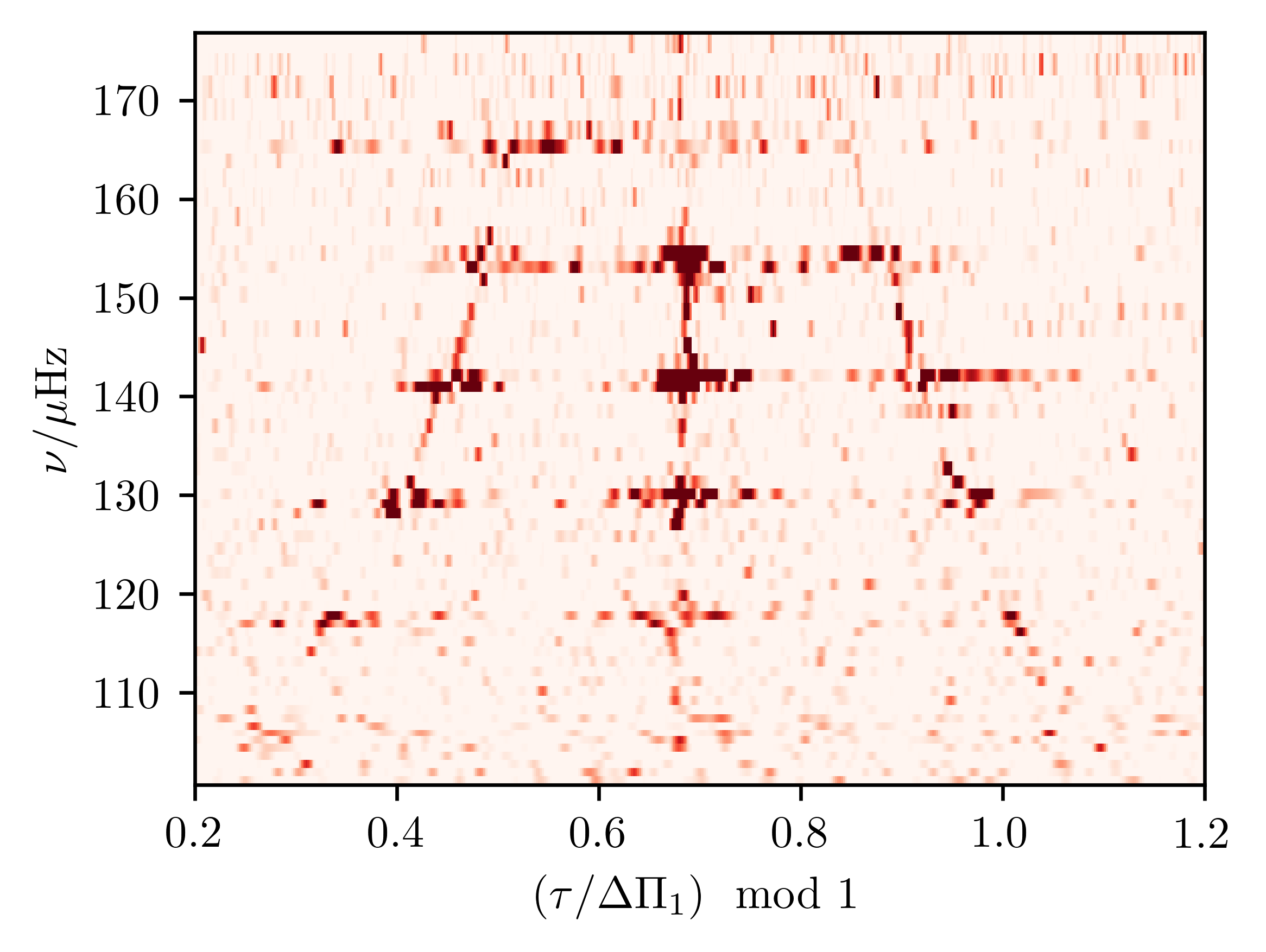}}{\node at (.9,.9){\textbf{(a)}};}
    \annotate{\includegraphics[width=.45\textwidth]{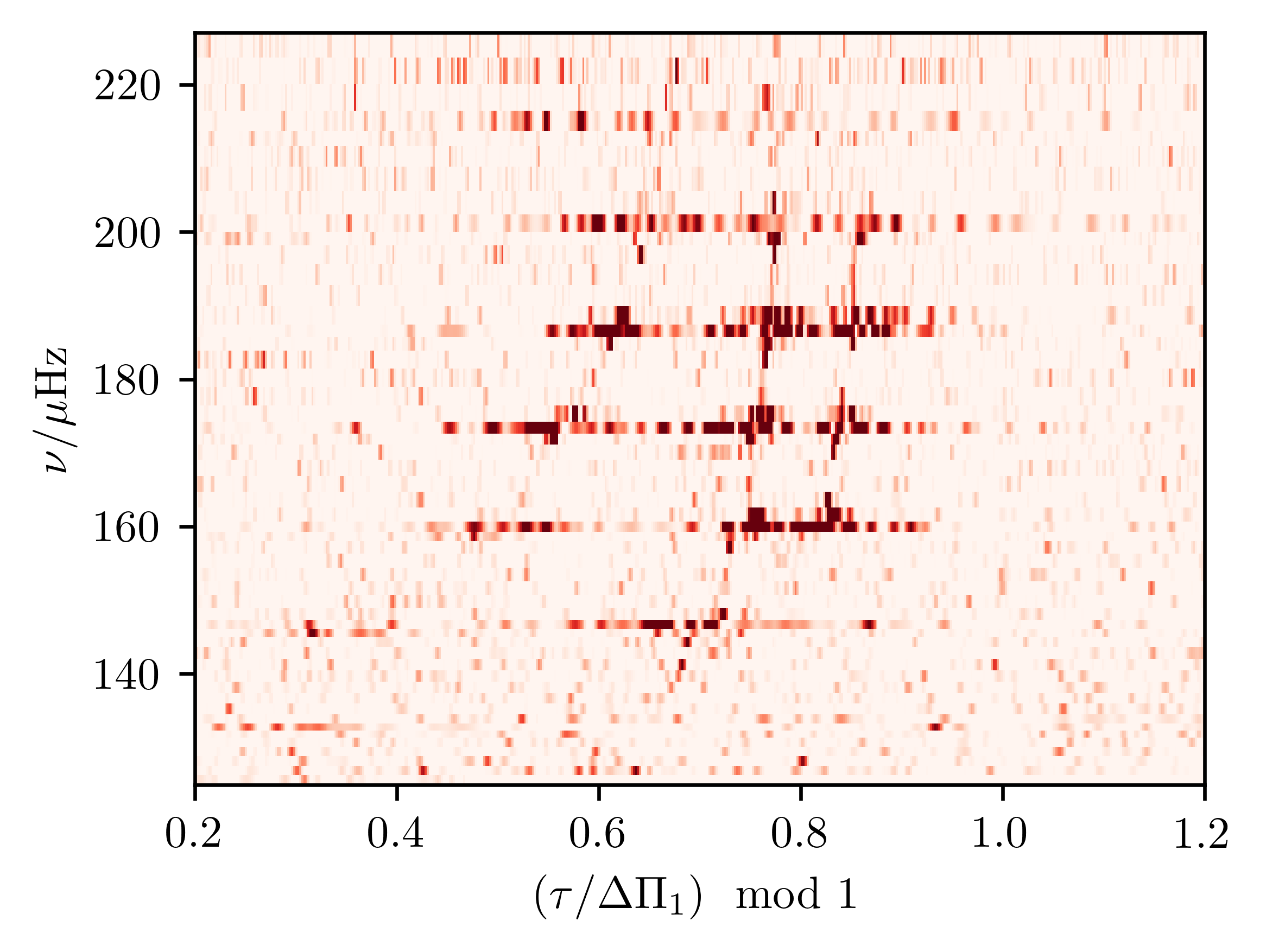}}{\node at (.9,.9){\textbf{(b)}};}
    \caption{Stretched \'echelle power diagrams for two red giants showing characteristic features of rotation and magnetism. \textbf{(a)}: KIC 10006097, showing symmetric rotational splitting. \textbf{(b)}: KIC 8684542, from the sample of \citet{li_magnetic_2022}, showing pronounced asymmetric rotational splitting indicative of core magnetism. The power spectrum indicates excess power along the g-mode ridges that do not correspond to identified and fitted modes in, e.g., their Fig. 3.}
    \label{fig:echelle}
\end{figure}
We then used the remaining S/N spectrum to estimate the posterior distribution of the $\ell = 1$ model parameters. At high radial order,pure $g$-modes approximately satisfy an asymptotic eigenvalue equation,
\begin{equation}
    1/\nu_{g} \approx  \Delta \Pi_{1}(n_{g} + \epsilon_{g}), 
    \label{eq:gmode}
\end{equation}

where $\Delta \Pi_{1}$ is the period spacing for $\ell = 1$ modes, $n_{g}$ is the g-mode radial order and $\epsilon_{g}$ is the phase offset. To calculate the frequencies of mixed-modes, these pure g-modes must be coupled to the pure p-modes. For this purpose we used the matrix construction of \cite{2021ApJ...920....8O} \citep[see also][]{2010Ap&SS.328..259D}. That is, mixed modes emerge as the eigenvalues ($\omega$) of the following, 

\begin{equation}
    \left(\begin{bmatrix}
    -\mathbf{\Omega}^2_p & \mathbf{L}\\
    \mathbf{L}^\dagger & -\mathbf{\Omega}^2_g
    \end{bmatrix} + \boldsymbol{\omega}^2 \begin{bmatrix}
    \mathbb{I} & \mathbf{D}\\
    \mathbf{D}^\dagger & \mathbb{I}
    \end{bmatrix}\right)\mathbf{v} = 0,
    \label{eq:coupling2}
\end{equation}
where $\boldsymbol{\Omega}_p=2\pi\boldsymbol{\nu}_p$ and $\boldsymbol{\Omega}_g=2\pi\boldsymbol{\nu}_g$ are diagonal matrices containing the angular frequencies of the pure p- and g-mode frequencies ($\boldsymbol{\nu}_p$ and $\boldsymbol{\nu}_g$), $\boldsymbol{v}$ is the eigenvector specifying mixed-modes as a combination of pure p- and g-modes, $\mathbf{L}$ and $\mathbf{D}$ are coupling matrices (with elements $L_{ij}$ and $D_{ij}$ respectively). In general, the elements of these coupling matrices vary with their associated p- and g-mode frequencies. We parameterised this frequency dependence as $L_{ij} \sim \omega_{g,j}^2 \cdot p_{L}$, where $p_{L}$ is a scalar, and similarly $D_{ij} \sim \omega_{g,j}/\omega_{p_i} \cdot p_{D}$, where $p_{D}$ is a scalar. A full motivation for this parameterisation will be provided in Nielsen et al. (in prep.). Therefore, for a given set of pure p- and g-mode frequencies, mixed mode frequencies can be described using the introduction of two parameters, $p_{L}$ and $p_{D}$. For each star we sample these as random, independent variables.

The parameters $\Delta \Pi_{1}$, $\epsilon_{g}$, $p_{L}$ and $p_{D}$ were used to provide unperturbed $m$ = 0 mode frequencies, to which we introduced the perturbations due to core rotation and a core magnetic field. In the presence of slow rotation, as is the case in these red giants, modes are perturbed linearly. Here, we approximate the rotation as happening in the core, ignoring the much slower envelope rotation \citep{2013A&A...549A..75G}. Under such assumptions, pure g-modes are perturbed according to, 

\begin{equation}
    \nu'_{m, g} = \overline{\nu_{g}} + m\dnu_{\mathrm{rot},g}, \label{eq:linearrot}
\end{equation}

where $\overline{\nu_{g}}$ is the unperturbed frequency and $\dnu_{\mathrm{rot},g}$ is the rotational splitting of the pure g-modes. In the following we will drop the subscript g for simplicity. 

A magnetic field in the core will also perturb the pure g-mode frequencies. In the following we used models consistent with those established in \cite{li_magnetic_2022}, which are subject to the constraint that effects of non-axisymmetry of the magnetic field are negligible. Unlike the case of rotation, a magnetic field will also impact the $m$=0 modes, such that we have, 

\begin{equation}
    \dnu_{\mathrm{mag}, m = 0} = (1- a)\delta\nu_{\mathrm{mag}}\bigg(\frac{\nu_{\mathrm{max}}}{\nu}\bigg)^3, 
    \label{eq:linearB1}
\end{equation}

\begin{equation}
    \dnu_{\mathrm{mag}, m = \pm 1} = (1+ a/2)\delta\nu_{\mathrm{mag}}\bigg(\frac{\nu_{\mathrm{max}}}{\nu}\bigg)^3,
    \label{eq:linearB2}
\end{equation}

where $a$ is a parameter dependent on the field topology, dependent on an average of the radial field strength weighted by a second order Legendre polynomial. As such, the value ranges between $-0.5\,<\,a\,<\,1$, with the maximum negative value corresponding to a field concentrated about the equator and maximum positive values corresponding to a field concentrated at the poles. A full inversion for the structure of the field is not possible, given the degeneracy between fields of different spatial scales \citep[see][]{2021MNRAS.504.3711L, 2023A&A...676L...9M}. The parameter $\delta \nu_{\mathrm{mag}}$ is dependent on an average of the radial field strength (see section \ref{sec:models}). It should be noted that the assumption of non-axisymmetry are met only when the ratio of the magnetic to the rotational splitting is less than one. 

Our total model for the perturbed pure g-mode frequencies is thus, 
\begin{equation}
    \nu'_{m} = \overline{\nu} + m\dnu_{\mathrm{rot}} +\dnu_{\mathrm{mag}, m} \label{eq:perturbed}.
\end{equation}
Given values for the parameters $\Delta \Pi_{1}$, $\epsilon_{g}$, $p_L$ and $p_D$, we may then calculate the resulting mixed-mode frequencies by performing mode-coupling calculations via equation \ref{eq:coupling2} for each $m$ separately, linearly perturbing the pure g-mode frequencies as above to describe core rotation and a core magnetic field. For this purpose, we neglect the effects of rotation and magnetism on the pure p-modes, which are largely insensitive to the core.

The final additional parameter required to describe the mode frequencies is the $\ell$ = 1 small frequency spacing ($\delta \nu_{01}$), which describes the deviation between the pure p-mode $\ell = 1$ frequencies and the midpoint of the adjacent $\ell$ = 0 mode frequencies. The complete set of parameters required to describe the frequencies numbers 8 ($\Delta \Pi_{1}$, $p_L$, $p_D$, $\epsilon_{\mathrm{g}}$, $\delta \nu_{01}$, $\dnumag$, $a$, $\dnurot$).

Similar to the analysis in section \ref{sec: 2,0}, we used these frequencies to fit a forward model to the power spectrum, described by a sum of Lorentzian profiles. Linewidths are fixed at the value of the $\ell = 0$ linewidth modified by the $\zeta$ function, $\Gamma_{\ell = 1} = \Gamma_{\ell = 0}(1 - \zeta)$. This accounts for the reduction in linewidth for g-dominated mixed modes, where the mode inertia is large. We assume mode heights can be approximated by the product of the envelope height from the $\ell = 0, 2$ model with the relative mode visibility of $\ell = 1$ modes (V$_{\ell=1}$), which for \textit{Kepler} is V$_{\ell =1}$ = 1.505 \citep{2012A&A...537A..30M, 2017ApJ...835..172L}. Additionally, the relative power between modes in a multiplet depends on stellar inclination, $i$. This is frequently accounted for by multiplying mode heights by the factor $\mathcal{E}_{\ell, |m|}(i)$. For dipole modes with $m$ = 0 this is given by $\mathcal{E}_{1, 0}(i)$ = $\cos^2(i)$, and for $m$ = $\pm$1 by $\mathcal{E}_{1, |1|}(i)$ = $1/2 \sin^2(i)$. We included this in our model, allowing $i$ to vary as a free parameter, setting a uniform prior in the range 0$\degree$ to 90$\degree$.

To estimate the posterior distribution on the parameters of the $\ell = 1$ model we use the \texttt{Dynesty} nested sampling package \citep{2020MNRAS.493.3132S}. This relies on establishing a set of prior distributions for each of the model parameters. 

\subsubsection{Priors on $\Delta \Pi_{1}$, $p_L$, $p_D$, $\epsilon_{\mathrm{g}}$, $\delta \nu_{01}$}\label{sec:priors gmodes}

To construct priors on these parameters, we exploited the so-called `stretched' \'echelle diagram construction \citep{2016A&A...588A..87V}. In the asymptotic approximation, mixed modes are the roots of the characteristic equation
\begin{equation}
    \tan\theta_{p}(\nu) \tan\theta_{g}(\nu) - q(\nu) = 0,\label{eq:mixed}
\end{equation}
where $q$ is a coupling strength (in most cases approximated as a constant), and $\theta_{p}$, $\theta_{g}$ are smooth functions of frequency constructed such that at pure p- and g-mode frequencies $\nu_p$ and $\nu_g$,
\begin{equation}
    \theta_{p}(\nu_p) = \pi n_p;\text{ and } \theta_{g}(\nu_g) = \pi n_g.\label{eq:theta}
\end{equation}
Given observational access to only mixed modes, but also inferences of notional p-mode frequencies, g-mode period spacings, and coupling strengths consistent with Eqs. (\ref{eq:gmode}),  (\ref{eq:mixed}) and (\ref{eq:theta}), one may invert Eq (\ref{eq:mixed}) to produce ``stretched'' frequencies $\nu_g$ associated with each mixed mode $\nu$. While several numerical formulations for doing this exist (e.g. \citealt{mosser_probing_2012,mosser_period_2015,mosser_period_2017,mosser_period_2018,gehan_core_2018,gehan_automated_2021}), \cite{ong_gehan_2023} prescribe an analytic expression,
\begin{equation}
    {1 \over \nu_g} \equiv \tau(\nu) \sim {1 \over \nu } + {\Delta\Pi_l \over \pi} \arctan \left(q \over \tan \theta_p\right),
\end{equation}
assuming that the pure p-modes are affected by neither rotation nor magnetism. Traditionally, these stretching functions are applied to mixed-mode frequencies fitted in advance from the power spectrum. In this work, we instead apply the stretching directly to the frequency coordinate of the power spectrum. Having done so, the morphology of the resulting stretched period-\'echelle power diagrams correspond directly to the linear expressions Eqs (\ref{eq:linearrot}) to (\ref{eq:linearB2}). 

To exploit this diagram to construct priors we note two features:
\begin{enumerate}
    \item If the correct values of $\Delta \Pi_{1}$ and $q$ are used to construct the diagram, modes of given azimuthal order should sit in distinct ridges. Therefore, by varying these parameters manually and identifying those which return the most well defined ridges we arrive at initial estimates of $\Delta \Pi_{1}$ and $q$.
    \item In the absence of a magnetic field, $m=0$ modes would align vertically in a ridge at the value of $\epsilon_{\mathrm{g}}$. As such, once we have settled on the combination $\Delta \Pi$ and $q$, we can use the central ridge as an estimate of $\epsilon_{\mathrm{g}}$. 
\end{enumerate}
 We varied these parameters by hand using the interactive tool introduced in \citet{reggae_RN}. We show examples of the resulting power diagrams in figure \ref{fig:echelle}. 
 
 Priors on $\Delta \Pi_{1}$ and $\epsilon_{\mathrm{g}}$ were set according to a normal distribution centered on our estimate from the stretched \'echelle. 
 Uncertainties on $\Delta \Pi_{1}$ from methods exploiting stretched \'echelles are on the order of a few percent \citep{2016A&A...588A..87V}. The only literature work using stretched \'echelles to measure $\epsilon_{\mathrm{g}}$ is \citet{mosser_period_2018}. There-in the mean uncertainty on $\epsilon_{\mathrm{g}}$ is $\approx$ 30\%. We also note that previous uncertainty estimates do not account for the presence of magnetic asymmetry. For cases where $m$ = 0 components have been significantly perturbed by a magnetic field, the combination of parameters constructing the most vertical $m=0$ ridge will not be an accurate representation of the true values. In an attempt to quantify this effect, we constructed a mock spectrum with values of $\dnumag$ and $a$ consistent with those reported for KIC8684542 by \citet{2023A&A...680A..26L} (the full set of asymptotic values used can be found in the appendix).
 We found that the difference between the injected and recovered $\Delta \Pi_{1}$ was below the 1\% level. The difference between the input $\epsilon_{\mathrm{g, \mathrm{input}}}$ and that from the hand tuned stretched \'echelle was more significant, at approximately 10\%, but remained below the mean uncertainty reported in \citet{mosser_period_2018}.
 
 We set the width of the prior on $\Delta \Pi_{1}$ 
 as 10\% of the mean. As previously discussed, the average fractional uncertainty on $\epsilon_{\mathrm{g}}$ reported in the literature is $\approx$ 30\% \citep{mosser_period_2018}. The computational expense associated with nested sampling scales with the volume of prior space, such that setting a very wide prior leads to the calculation becoming infeasible. As such, the width of the prior on  $\epsilon_{\mathrm{g}}$ was set to 30\% of the mean. To ensure our results are not prior dominated, we visually inspected the posterior versus prior distributions on $\epsilon_{\mathrm{g}}$.

Exploiting the stretched period \'echelles required us to use the asymptotic expression for mixed modes, rather than the matrix formalism used in the sampling. Following \citet{ong_gehan_2023} the value of $q$ evaluated at $\numax$ can be determined from the matrix coupling parameters as
 \begin{equation}\label{eq: q}
     q \approx \frac{1}{\Delta \nu \Delta \Pi_{1}} \left(L + \omega^2 D \over 8 \pi \nu^2\right)^2 = \frac{1}{\Delta \nu \Delta \Pi_{1}} \left({\pi \over 2 }(p_L + p_D)\right)^2,
 \end{equation}
with all frequencies evaluated at $\nu_\text{max}$. This expression indicates that, from a single value of $q$ alone, it is not possible to uniquely identify $p_L$ and $p_D$. However, given they are informed by the internal profile of the star, certain values will be more physically motivated. To identify characteristic values of $p_L$ and $p_D$ for our stars we exploited the grid of stellar models used in \citet{2021ApJ...922...18O}. For a given model star in this grid, values of $p_{L}$ and $p_{D}$ were subsequently calculated from the corresponding coupling matrices. Stellar tracks were calculated with masses from 0.8 M$_{\odot}$ to 2 M$_{\odot}$, and [Fe/H] from $-1.0$ dex to $0.5$ dex.
 
For a given star, we select stellar model tracks in a mass range consistent with those reported in Y18. We then located models with $\numax$ in the range $\nu_{\mathrm{max, obs}}\pm \sigma(\nu_{\mathrm{max, obs}})$, $\Delta \Pi_{1}$ in the range $\Delta \Pi_{1, \mathrm{prior}} \pm  0.1\Delta \Pi_{1, \mathrm{prior}}$ and a value of $q$ consistent with that derived from the stretched \'echelle (to within 10\%). The means of the distributions of $p_L$ and $p_D$ in the selected models were taken as the means of the normal distributions we used as priors on $p_L$ and $p_D$. For all the stars in our sample, the model grid returned values of $p_L$ and $p_D$ in ranges spanning standard deviations that were on average 2\% and 10\% of the mean model values, respectively. Accordingly we set the widths of the priors on $p_L$ and $p_D$ for each star at 10\% of their mean value.

\subsubsection{Priors on $\dnumag$, $a$, $\dnurot$}
\begin{figure}
    \centering
    \includegraphics[width=1\linewidth]{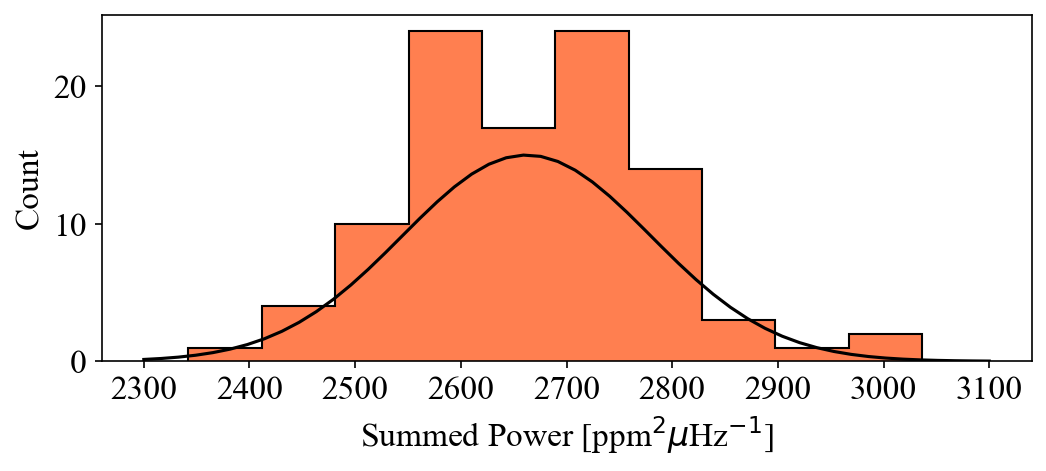}
    \caption{Distribution of summed power across ridges defined by $\dnurot$ = 0, $\dnumag$ = 0 and $a$ = -0.5 for a white noise spectrum stretched according to the asymptotic parameters of KIC10006097. The number of realisations has been increased from 50 to 100 for illustrative purposes.}
    \label{fig:summed power distribution}
\end{figure}

\begin{figure}
    \centering
    \includegraphics[width=0.9\linewidth]{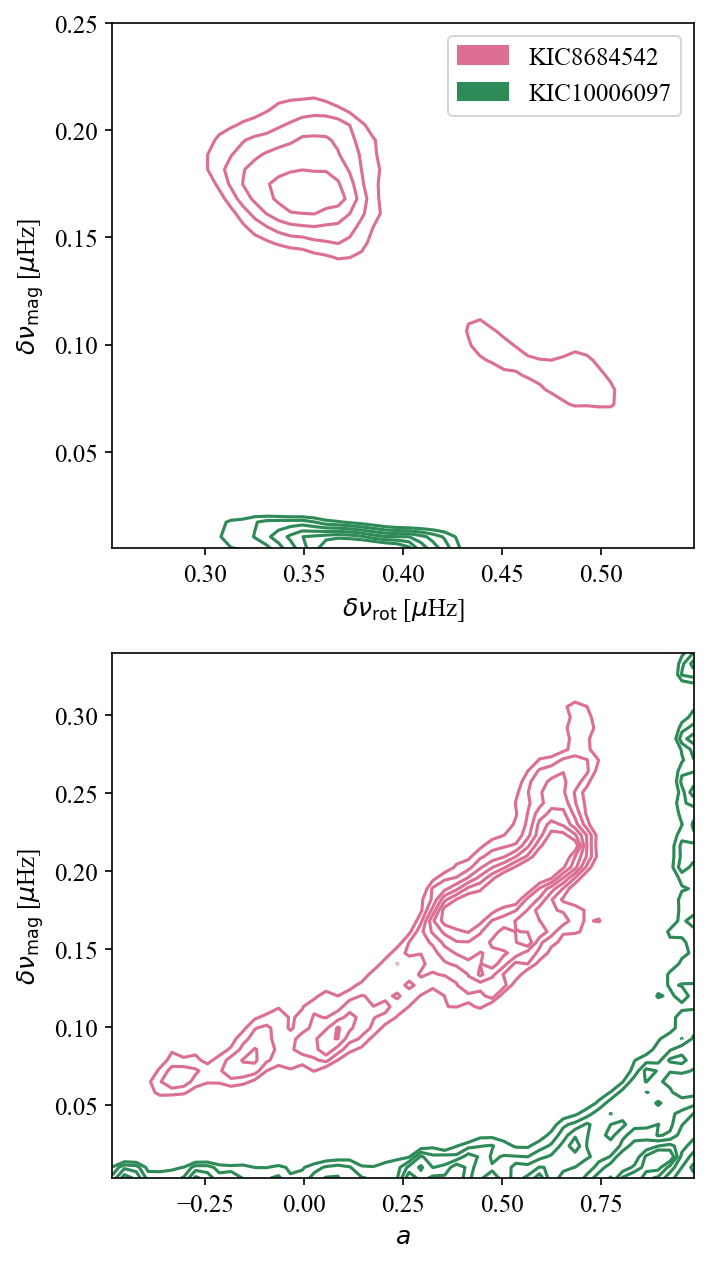}
    \caption{Top panel: Inverse of the 2-dimensional H0 likelihood space for $\dnumag$ and $\dnurot$. Two stars are shown, KIC8684542 and KIC10006097. Bottom panel:  Inverse of the 2-dimensional H0 likelihood space for $\dnumag$ and $a$ for KIC8684542 and KIC10006097.}
    \label{fig:likelihood dnu}
\end{figure}

Once $\Delta \Pi_{1}$, $q$ and $\epsilon_{g}$ have been set, ridges in the stretched \'echelle can be approximated using the three remaining parameters in our mixed mode model. Therefore, for each star we constructed a grid of templates describing possible ridges given test values of $\dnumag$, $a$ and $\dnurot$. We uniformly sample $\dnumag$ in the range 0 to 0.2$\mu$Hz, $\dnurot$ in the range 0.0 to 0.8$\mu$Hz and $a$ from -0.5 to 1. Our grid had 50 points in each direction, such that the resolution on the magnetic splitting is 0.004$\mu$Hz, on the rotation it is 0.02$\mu$Hz and on $a$ is 0.03. 

To establish the parameter values required to best describe the data we performed a null hypothesis test (H0 test). For a given star, we summed the total observed power in these ridges, and established the likelihood that we would observe the resulting power just due to white noise (the H0 likelihood). If we were summing power in N bins without performing the stretching, this would simply amount to the likelihood of drawing a given value of summed power from a $\chi^2$ distribution with 2N degrees of freedom. However, the stretching introduces correlation between bins, such that the number of degrees of freedom in the stretched spectrum is no longer 2N. An analytical definition of the correct number of degrees of freedom required to describe a stretched spectrum is yet to be established. Given the degree of stretching depends on $\DPi_{1}$ and $q$, this will vary on a star-by-star basis. 

To approximate the statistics for the summed stretched power we therefore used white noise simulations. For each target we performed 50 realisations of a white noise spectrum evaluated on the same frequency grid as the real data. We then stretched this spectrum according to the value of $\DPi_{1}$ and $q$ used in the prior and calculated the summed power about the predicted ridges according to our grid of $\dnumag$, $a$ and $\dnurot$ (as would be done for a real star). This resulted in 50 realisations of a 3 dimensional summed power array for a given star. Given each point corresponds to the sum of a large number of $\chi^{2}$ distributed parameters, the resulting sum should be distributed according to a Gaussian (according to the central limit theorem). Accordingly, we calculated the H0 likelihood of the real data for a combination of $\dnumag$, a, $\dnurot$ via, 

\begin{equation}
    \mathcal{L}(\Theta|H0) \approx \mathcal{N}(\mu_{\Theta,WN}, \sigma_{\Theta, WN}),
\end{equation}
where $\mu_{\Theta,WN}$ and $\sigma_{\Theta, WN}$ are the mean and standard deviation of the white noise realisations. The parameters are $\Theta$ = ($\dnumag$, $a$, $\dnurot$). Figure \ref{fig:summed power distribution} shows the distribution of summed power in a single cell of the 3-d array for a white noise spectrum stretched according to the asymptotic parameters of KIC10006097.

The width over which we summed the power about the predicted ridge was informed by the expected (stretched) line-width for g-dominated modes. At the base of the RGB, the distribution of radial mode linewidths peaks at $\approx$ 0.15$\mu$Hz \citep{yu_asteroseismology_2018}. The dipole mode linewidth for a given mixed mode then scales as $\Gamma_{1}(\nu) = \Gamma_{0}(1 - \zeta(\nu))$. For a mode $\nu_{i}$ with $\zeta(\nu_{i}) = 0.9$, this implies $\Gamma_{1}(\nu)$ $\approx$ 0.015$\mu Hz$. Therefore, we sum power in a width of 0.03$\mu$Hz. Given this definition was set using an arbitrary selection of $\zeta$, we tested 10 different widths up to a maximum of 0.15$\mu$Hz. For KIC10006097 we found the resulting value of $\dnurot$ was consistent across widths from 0.015 to 0.084 $\mu$Hz. At widths larger than 0.084 $\mu$Hz, the measured value of $\dnurot$ was consistently smaller by $\approx$ 0.1$\mu$Hz. 

Examples of the 2-d distributions in likelihood for the \'echelle diagrams shown in figure \ref{fig:echelle} can be seen in figure \ref{fig:likelihood dnu}. The H0 likelihood was then marginalised over each axis, and the minima of the 1D distributions used to inform the mean of the prior used in the sampling.

In a handful of cases the likelihood space was multimodal. To identify which mode best described the data, we manually vetted the associated ridges and subsequently reduced the range to exclude the spurious peak. This multimodality was a by-product of the use of H0 likelihood, as a model that is not necessarily the best descriptor of the signal can still capture power that is very unlikely to be the result of noise (for example residual power from $\ell = 0, 2$ modes).  

\begin{figure}
    \centering
    \includegraphics[width=1\linewidth]{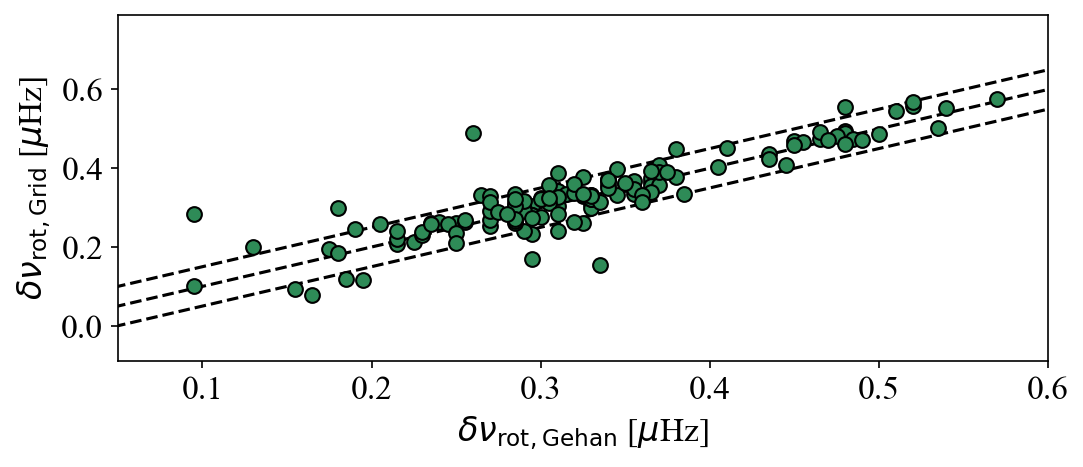}
    \caption{Core rotational splitting from the template matching technique versus those reported in G18. Black dotted lines show the 1-1 relation $\pm$ 0.05 $\mu$Hz}
    \label{fig:rotation priors}
\end{figure}

Given these likelihoods are conditional on the combination of $\Delta \Pi_{1}$, $q$ and $\epsilon_{g}$, we did not use the width of the minima in the H0 likelihood to establish the width of the prior. To establish the most appropriate width to set on the prior on rotational splitting, we compared our values to those from \cite{gehan_core_2018} (henceforth G18). The resulting differences in measured rotation rates give a better estimate of the uncertainty associated with varying asymptotic parameters. Of the 334 stars in our target list, 142 are also in G18. The differences between the values in that catalogue and those we measured can be well approximated using a normal distribution with $\mu$ = 0.00 $\mu$Hz and $\sigma$ = 0.05 $\mu$Hz (see figure \ref{fig:rotation priors}). As such our prior was $\mathcal{N}(\dnu_{\mathrm{rot, echelle}}, 0.05 \mu \mathrm{Hz})$.

\begin{figure}
    \centering
    \includegraphics[width=1\linewidth]{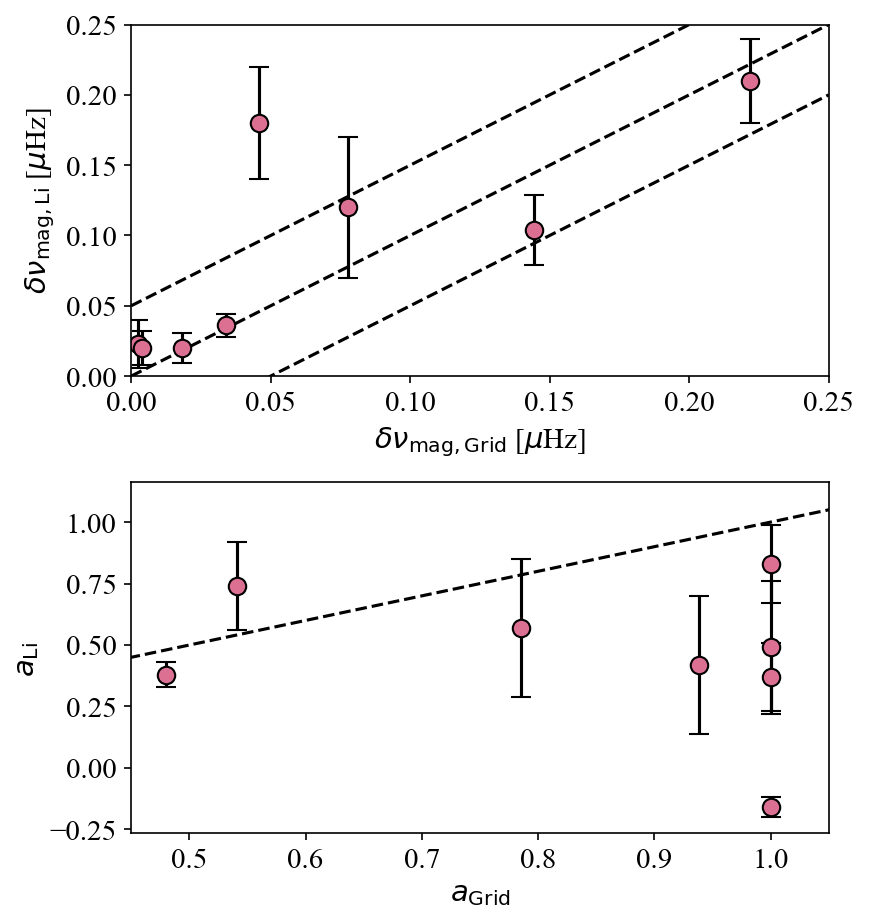}
    \caption{Top panel: Magnetic splitting, $\dnumag$, measured using summed power versus that reported in L23. Black dotted lines are the 1-1 relation $\pm$ 0.05$\mu$Hz. Bottom panel: Topology parameter, $a$, measured using summed power versus that reported in L23. Black dotted line is the 1-1 relation.}
    \label{fig:magnetic priors}
\end{figure}

The largest catalogue of magnetic parameters is that of \cite{2023A&A...680A..26L} \citep[henceforth L23, see also][henceforth L22]{li_magnetic_2022}. Of the 13 stars listed there, 8 appear in our target list. We found our values of $a$ differed substantially (see figure \ref{fig:magnetic priors}), with a preference for large values, which could be a consequence of the correlation with $\Delta \Pi_{1}$ (noted in L23). As such we ignore the result from the summed stretched power and set a uniform prior on $a$ between -0.5 and 1 for all stars. 

Our values of $\dnumag$ are in better agreement (see figure \ref{fig:magnetic priors}), with a standard deviation of 0.05$\mu$Hz. We set the prior as uniform, allowing values in the range [$\dnu_{\mathrm{mag, echelle}}$ - 0.15$\mu$Hz, $\dnu_{\mathrm{mag, echelle}}$ + 0.15$\mu$Hz]. For cases where this width would cause the prior to allow negative values of $\dnumag$, we set the lower limit on the prior at 0. In one outlier, the grid method has a $\dnumag$ which is smaller than that in L23 by 0.13$\mu$Hz. This is likely a consequence of selecting the value of $\Delta \Pi_{1}$ that made the $m$ = 0 ridge appear most vertical (see section \ref{sec: methods}). Given setting a wider prior on $\dnumag$ for all stars would result in significant additional computational expense, we manually vetted posteriors and best fit models and only expanded the prior ranges where necessary. This amounted to expanding the prior range on $\dnumag$ for KIC8684542.

%Describe method of summing power, how we grid (and decide on range to grid over), fitting mean magnetic field strength  

\section{Results}\label{sec:results}
Of the 334 targets we report the magnetic and rotational parameters in 302 stars. Those 32 stars for which we do not report results are cases where the posterior distribution was simply a replica of the prior. Such were the result of either low SNR or low inclination such that the rotational splitting was not well constrained. Example corner plots for two stars drawn from the sample of 302 (for which we report results) are shown in the appendix (figures \ref{fig:11515377 corner} and \ref{fig:7018212 corner}). For a comparison of our values of $\Delta \Pi_{1}$, $q$ and literature values, see appendix \ref{appendix lit}. The full catalogue of asymptotic g-mode parameters, alongside $\dnurot$, $\dnumag$ and $a$ is available online, with the first 10 rows presented in Table \ref{tab:example_of_results}.

\subsection{Rotational splitting}
\begin{figure}
    \centering
    \includegraphics[width=1\linewidth]{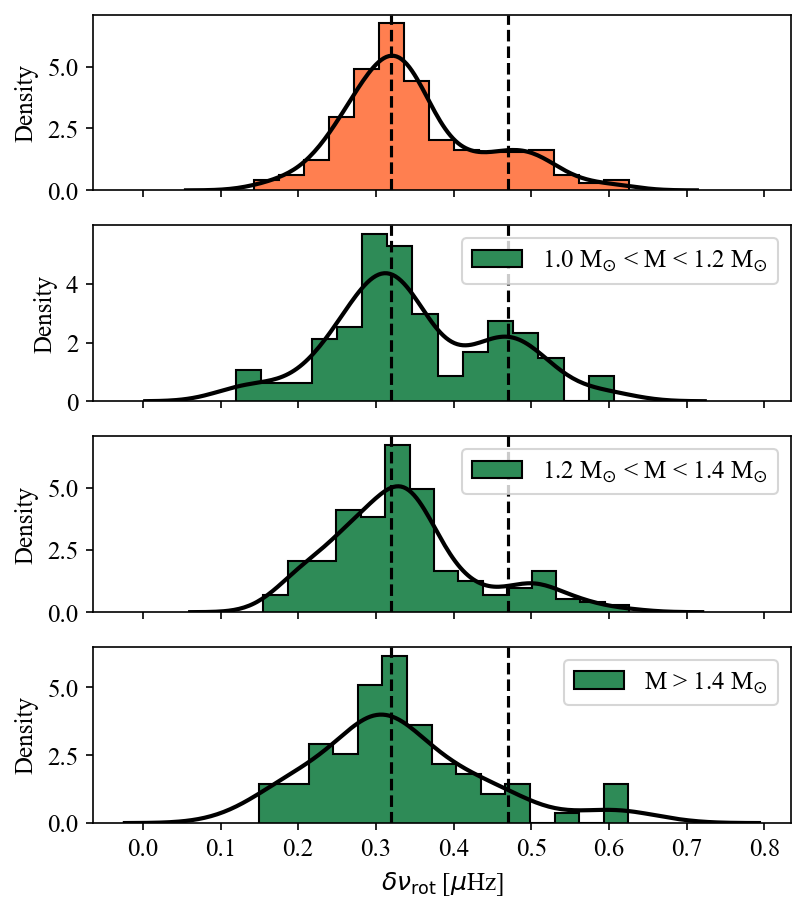}
    \caption{Distribution of $\dnurot$. The top panel shows the distribution for the measurements made in this work. The following three panels include values from G18 with $\numax$ > 100$\mu$Hz and are separated into three mass ranges. The black curves show KDEs of the distributions. Black dotted lines mark the locations of the two peaks identified in the lowest mass set to guide the eye.}
    \label{fig:3 mass}
\end{figure}

The distribution of core rotational splitting is shown in the top panel of figure \ref{fig:3 mass}, and is bimodal. The more populous peak is located at $\approx$ 0.32$\mu$Hz with the secondary peak at $\approx$ 0.47$\mu$Hz. There does not appear to be any strong correlation between the rotational splitting (or associated bimodality) and the remaining asymptotic parameters (see figure \ref{fig:obs corner}, which shows a corner plot of the distribution of asymptotic parameters across the whole population). Notably the distributions of $\dnumag$ and $a$ with $\dnurot$ > 0.4$\mu$Hz are consistent with those in the remaining catalogue.

\begin{figure}
    \centering
    \includegraphics[width=1\linewidth]{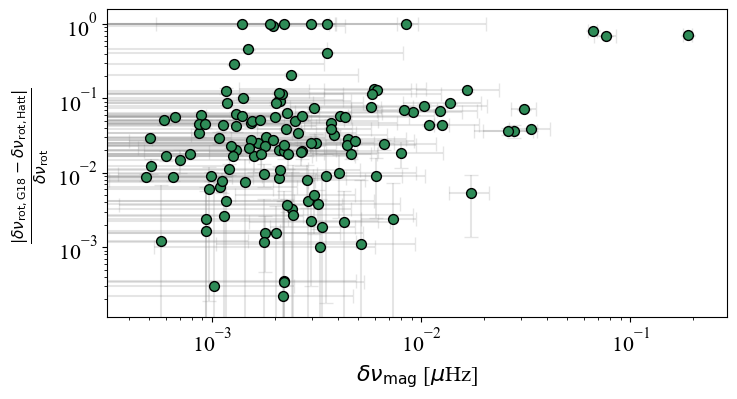}
    \caption{Fractional difference in $\dnurot$ measured here and reported in G18 as a function of $\dnumag$.}
    \label{fig:diff dnurot}
\end{figure}

Revisiting the 142 stars which appear in G18 which we used to inform the width of the prior on $\dnurot$, we found general agreement, with $\approx$ 80\% agreeing to 10\% or better. As can be seen in figure \ref{fig:diff dnurot}, we found a correlation between the difference in $\dnurot$ with $\dnumag$. For the star with the largest value of $\dnumag$ (KIC8684542), the difference in $\dnurot$ is of order 0.23 $\mu$Hz. This was also noted in L23. Given the small values of magnetic splitting present in the vast majority of our targets, the offset in $\dnurot$ is below the scale of the uncertainties. As such, this is unlikely to impact previous conclusions regarding the distribution of rotational splittings on a population scale. We note a small number of stars (9) with differences in $\dnurot$ exceeding 0.1 despite having $\dnumag$ values below 0.001. This is a symptom of low SNR on the $m = \pm 1$ components of multiplets due to low stellar inclination, the average in the 9 cases being 45$\degree$.

\subsection{Magnetic Parameters}
\begin{figure}
    \centering
    \includegraphics[width=1\linewidth]{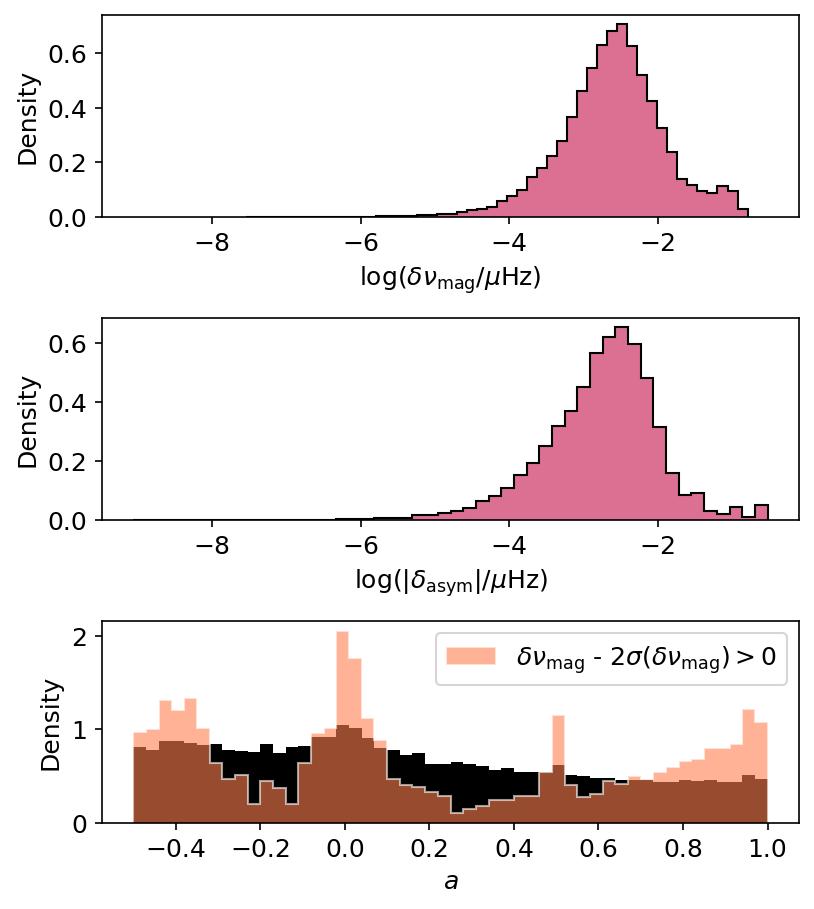}
    \caption{Distributions of magnetic parameters. Histograms are built using 100 draws from the posterior for each of the 302 stars. Top panel: Distribution of $\dnumag$. Middle panel: Distribution of $\delta_{\mathrm{asym}}$. Bottom panel: Distribution of $a$. }
    \label{fig:mag hist}
\end{figure}

The distribution in magnetic splittings peaks about log$_{10}$($\dnumag$) = -2.54, with a standard deviation of $\sigma$(log$_{10}$($\dnumag$)) = 0.45 (see figure \ref{fig:mag hist}). This indicates that, from an observational perspective, perturbations to mode frequencies due to a core magnetic field of the scales reported in L22 and L23 are uncommon regardless of asymmetry. In total we find the mean on the posterior of the magnetic splitting is at least 2$\sigma$ from zero in 23 stars, approximately 8\% of the total sample. The corner plots and stretched \'echelle diagrams for these 23 stars are available as online materials.

The measurements of the topology parameter $a$ span the full range in values allowed by the prior, with a peak at zero - the value which minimizes the asymmetry (see figure \ref{fig:mag hist}). For stars in which the mean value of the posterior on $\dnumag$ is at least 2$\sigma$ from zero, 35\% have values of $a$ exceeding 0.5. These are inconsistent with a dipolar field, and must be identified with an architecture having the field concentrated more towards the poles than the equator. Values of $a$ below -0.2 occur in 30\% of the stars with significant $\dnumag$. These also cannot be the result of a dipolar field, instead being consistent with a field concentrated near the equator.

The asymmetry between modes in a multiplet is often used as an identifier of the presence of a magnetic perturbation. L22 and L23 quantify this using an additional parameter, $\delta_{\mathrm{asym}}$ = $3a\dnumag$. In our catalogue $\delta_{\mathrm{asym}}$ also peaks near zero, the distribution having a mean value log$_{10}$($\delta_{\mathrm{asym}}$)=$-$2.78 with a standard deviation of $\sigma$(log$_{10}$($\delta_{\mathrm{asym}}$)) = 0.67 (see figure \ref{fig:mag hist}). Only 12 stars have asymmetry parameters that are at least 2$\sigma$ from zero, making them easily identifiable by eye. All 8 of the stars that we have in common with L23 appear in this set. The remaining targets identified as having significant $\dnumag$ values but little asymmetry have not been previously identified. 

We find no obvious correlation between $\dnumag$ and any other asymptotic parameter, including the topology. This was also noted in L23. \cite{2022A&A...667A..68B} noted that not accounting for a magnetic perturbation could produce a systematic underestimate of $\Delta \Pi_{1}$ when using the techniques presented in \cite{2016A&A...588A..87V}. They simulated a star with a magnetic splitting of $\delta \nu_{\mathrm{mag}}$ = 0.4$\mu$Hz, which is much larger than those reported in L22 and L23 (where-in the maximum reported value is 0.21$\mu$Hz), and found this would cause a 1\% difference in the measured period spacing. We do not find a clear trend between the difference in our period spacing and that recorded in \cite{2016A&A...588A..87V} with the magnetic splitting. As such, for stars with the magnitudes of magnetic perturbation reported here, discrepant $\Delta \Pi_{1}$ alone cannot be used as an identifier.

For the 8 stars we have in common with \citet{2023A&A...680A..26L}, our values of $\dnumag$ are in broad agreement, with the mean absolute difference being 1.09$\sigma$. Agreement on $a$ is slightly worse, with a mean absolute difference of 1.5$\sigma$. Our methods do differ, \citet{2023A&A...680A..26L} fit the asymptotic expression to mode frequencies rather than directly fitting a forward model to the spectrum. Additionally we differ in our method of coupling modes, where we used the matrix construction discussed in \citet{2021ApJ...920....8O} and \citet{2010Ap&SS.328..259D}, \citet{2023A&A...680A..26L} use the JWKB construction of \citet{1989nos..book.....U}. Finally, our priors on asymptotic parameters differ (see \citealt{2023A&A...680A..26L} for details).

\section{Discussion}\label{sec:discuss}

\subsection{Rotation and stellar mass}
As previously noted, the distribution of core rotational splitting appears bimodal. To better sample the underlying distribution, we expand our data to include stars from the G18 catalogue with $\numax$ in the range of the stars in our sample. To that end we selected stars from G18 that both do not appear in our target list and have $\numax$ greater than the minimum in our targets at $\nu_{\mathrm{max, min}}$ = 135$\mu$Hz. In total the expanded sample numbers 583 stars. Although it was not reported in G18, a bimodality in core rotation does appear to be present in their measurements for stars in this range in $\numax$ (see the discussion in Appendix \ref{appendix:bimodality}).

\begin{figure*}
    \centering
    \includegraphics[width=1\linewidth]{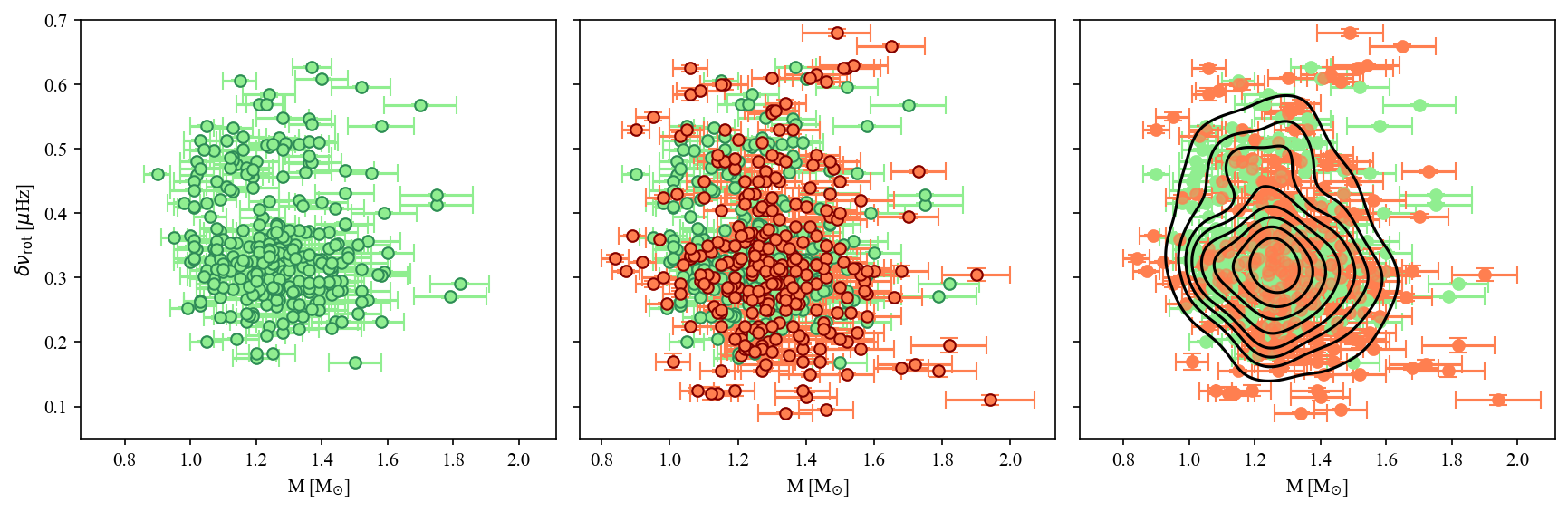}
    \caption{Distribution of $\dnurot$ as a function of stellar mass for the combined set of stars reported here and in G18. The left-most panel shows just the measurements reported in this work. In the center panel we fold in measurements from G18 (orange). The Black contours in the right hand panel are a kernel density estimate (KDE) highlighting the bimodality. There-in the bin-width has been set according to Scott's rule of thumb \citep{1992mde..book.....S}.}
    \label{fig:dnurot mass}
\end{figure*}

We show the distribution of $\dnurot$ as a function of stellar mass in figure \ref{fig:dnurot mass}. Masses are from Y18, where-in they are calculated via scaling relations with $\numax$, $\Dnu$ and $T_{\mathrm{eff}}$. In the left-most panel (showing the new measurements reported here without the additional stars from G18) it appears the divide between the two populations in rotation is mass dependent, with the more rapidly rotating peak preferentially populated with less massive stars. To identify whether the two apparently distinct over-densities remain in the sample with the additional stars from G18, a KDE estimate is plotted in black in the final panel in \ref{fig:dnurot mass}. There-in we can identify the clear over-density at $\dnurot$ $\approx$  0.3$\mu$Hz, alongside the less populous peak in density at $\dnurot$ $\approx$  0.5$\mu$Hz.

We divided the combined catalogue into stars in ranges 1.0M$_{\odot}$ < M$_{*}$ < 1.2M$_{\odot}$,  1.2M$_{\odot}$ < M$_{*}$ < 1.4M$_{\odot}$ and M > 1.4M$_{\odot}$. The resulting distributions in $\dnurot$ can be seen in figure \ref{fig:3 mass}. In the lowest mass range, the secondary peak is identifiable at $\dnurot$ 0.47 $\mu$Hz. In the mid range, the peak shifts upwards in $\dnurot$ to 0.50 $\mu$Hz and is less well populated. The set with the highest masses contains the fewest total stars and it is unclear whether a secondary peak is present. In total 25\% of the population have $\dnurot$ > 0.40$\mu$Hz.  

\begin{figure*}
    \centering
    \includegraphics[width=1\linewidth]{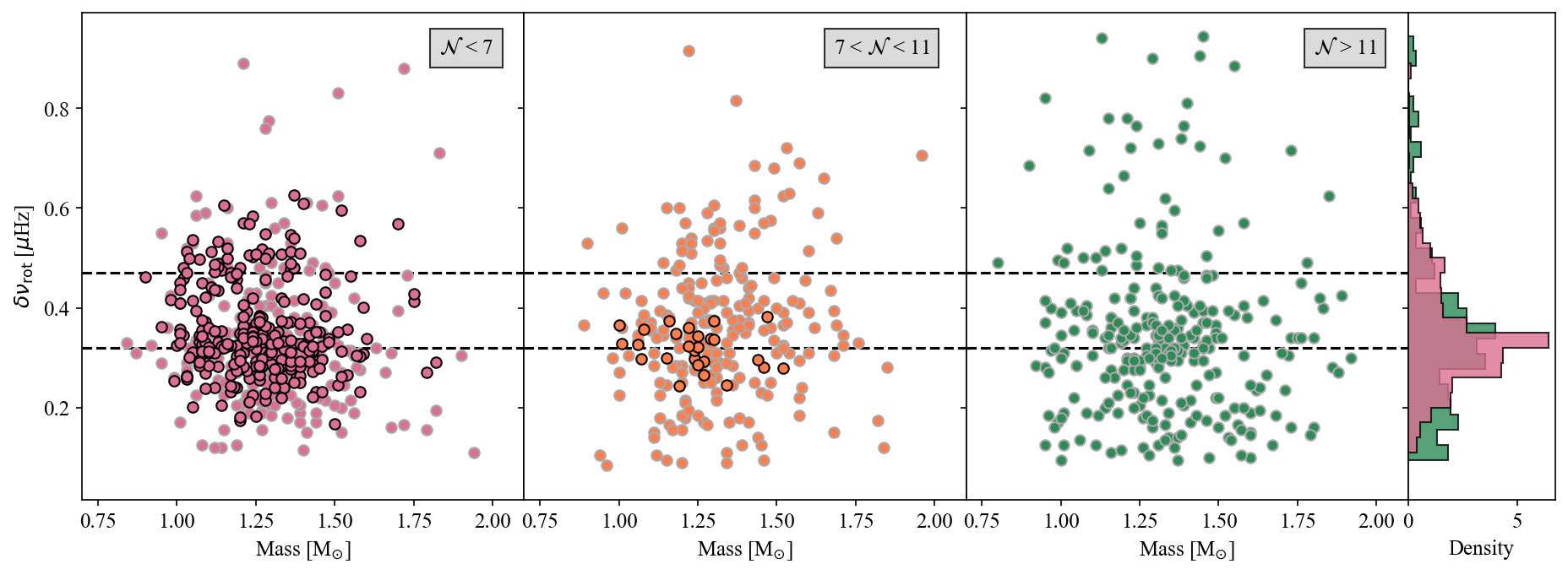}
    \caption{Distribution of $\dnurot$ reported in G18 (grey edgecolor) and here (black edgecolor) for stars in 3 different $\mathcal{N}$ ranges. The left-most panel shows targets with $\mathcal{N}$ < 7, the middle panel has targets in the range 7 < $\mathcal{N}$ < 11. The right-most shows stars with $\mathcal{N}$ > 11. Black dotted lines are at $\dnurot$ = 0.32 $\mu$Hz and 0.47 $\mu$Hz. Histograms on the right show the distribution in the stars with $\mathcal{N}$ < 7 in pink and with $\mathcal{N}$>11 in green.}
    \label{fig:numax dnurot}
\end{figure*}

In figure \ref{fig:numax dnurot} we show our measurements of $\dnurot$ and those from G18 as a function of stellar mass in three ranges of $\mathcal{N}$. The left-hand panel is a reproduction of figure \ref{fig:dnurot mass}. The following panels show this distribution with increasing mixed mode density, $\mathcal{N}$. As such, for a given mass, stars go from least to most evolved from the left-most to the right-most panel. G18 and \cite{mosser_probing_2012} concluded that core rotation rates in red giants decrease slightly as stars evolve, but do not undergo significant change. Indeed, the highest density of stars is at $\approx$ 0.3$\mu$Hz in all three subsets. However, the spread in the distribution increases, such that the secondary peak identified at 0.47$\mu$Hz appears to migrate to larger values with increasing $\mathcal{N}$. For stars with $\mathcal{N}$ < 7, 2.4\% of the population have $\dnurot$ > 0.6$\mu$Hz, this increases to 6.1\% for stars with $\mathcal{N}$ > 11. 

Red Clump stars are expected to have core rotation rates that differ from those on the RGB. However, given the cores are undergoing expansion due to the onset of helium burning, rotation rates are expected to decrease. This was observed to be the case in \cite{mosser_probing_2012}. There-in the authors reported splittings ranging from 0.01 to 0.1$\mu$Hz. Therefore, it is unlikely the subset of rapid rotators are misclassified clump stars.

One interpretation of this result is that the stars in the more rapidly rotating population in each range in $\mathcal{N}$ belong to a single population with a weaker rotational coupling between core and envelope, such that the efficiency of angular momentum transport has been reduced. Accordingly their cores would be able to more effectively spin up as they contract. As the core contracts, the envelope is undergoing expansion, such that we can use the evolution of core rotation with stellar radius to infer the sign of the dependence on core contraction. \citet{2014ApJ...788...93C} found that the increase in core rotation should scale with stellar radius as $\Omega_{\mathrm{core}} \propto R_{*}^{\alpha}$, with $\alpha$ taking a value of 1.32 or 0.58, depending on whether they include just rotational instabilities or fold in those due to magnetic torques in radiative regions. That is, the cores should spin up as the star evolves. This is in clear disagreement with the core rotation rates measured by \citet{mosser_probing_2012}, who found exponents of $-$0.5 for stars on the RGB and $-$1.3 in the RC. The mean values of stellar radius in the stars with $\mathcal{N}$ < 7 and $\mathcal{N}$ > 11 are 4.89 R$_{\odot}$ and 7.14 R$_{\odot}$ respectively. For representative stars at these radii to spin up from $\approx$ 0.5$\mu$Hz to $\approx$ 0.7$\mu$Hz would imply a relation of the form $\Omega_{\mathrm{core}} \propto R_{*}^{0.8}$. This exponent sits in between those predicted by \citet{2014ApJ...788...93C}.

\subsection{Magnetic Perturbations and Stellar Properties}
As noted in section \ref{introduction}, a core magnetic field could impact the transport of angular momentum within a star.  We found no clear correlation between the bimodality in $\dnurot$ and $\dnumag$ or $a$. However, given a strong core magnetic field is frequently hypothesised as a solution to the discrepancy between modelled and observed core rotation rates, constraints on the average field strength in a large sample of stars remain in demand. In the following sections we invert the observed magnetic splitting to constrain the average core magnetic field strength.

\subsubsection{Stellar Models}\label{sec:models}
According to L22, the root-mean-square (rms) of the radial field strength scales with the magnetic splitting as

\begin{equation}\label{eq:dnu_mag_B}
    \langle \mathrm{B}_{\mathrm{r}}^2 \rangle = \int_{r_i}^{r_o}K(r)\overline{B_{r}^2}dr =\frac{16 \pi^4\mu_{0}\numax^3\dnumag}{\mathcal{I}},
\end{equation}
where $\mathcal{I}$ is a factor determined by the internal structure of the star. This is given by, 
\begin{equation}
    \mathcal{I} = \frac{\int_{r_{i}}^{r_{o}}\big(\frac{N}{r}\big)^3\frac{dr}{\rho}}{\int_{r_{i}}^{r_{o}}\frac{N}{r}dr},
\end{equation}
where $N$ is the Brunt-Väisällä frequency, $r$ is radius and $\rho$ the stellar density. Therefore, we require models of the internal profile of our targets to invert our measured $\dnumag$ to give an estimate of the radial field strength.

To that end, we used Modules for Experiments in Stellar Astrophysics  \citep[MESA,][]{2011ApJS..192....3P, 2013ApJS..208....4P} to calculate a grid of stellar models with varying mass and metallicity. We calculated stellar evolution tracks with masses varying in the range of those reported in Y18 for our targets, spanning from 1M$_{\odot}$ to 2M$_{\odot}$ in increments of 0.05M$_{\odot}$. Metallicities ranged from -1.0dex to +1.0dex with a spacing of 0.25dex. We used a mixing length of $\alpha_{\mathrm{MLT}}$ = 2.29, which was found via calibration of a solar model. Overshoot was treated using the exponential formalism with $f_{1}$ = 0.015 and $f_{0}$ = 0.004. Mode frequencies were then calculated using GYRE. To avoid too much computational expense, we restricted the grid to just radial modes. 

A suitable model from this grid was selected for each star using the AIMS package \citep[Asteroseismic Inference on a Massive Scale,][]{2019MNRAS.484..771R}. To do this AIMS searches the grid for the region with the highest posterior probability given the observed parameters, then explores the surrounding space using an MCMC sampler \citep[\texttt{emcee},][]{Foreman-Mackey_2013}. For general use in fitting global properties, AIMS interpolates between grid points. However, we require internal profiles and are, therefore, restricted to selecting models in the grid. We choose the model with the highest posterior probability, but note this is naturally restricted by the grid resolution. AIMS provides several methods to apply surface corrections to model frequencies, of which we selected the method of \citet{2014A&A...568A.123B}. We provided $T_{\mathrm{eff}}$, [Fe/H], $\log(g)$, mass (from Y18) alongside $\Delta \Pi_{1}$, $\numax$, $\Dnu$ and radial mode frequencies from the $\ell = 2, 0$ model (see section \ref{sec: 2,0}) as observables. 

The only model parameter which could induce additional uncertainty in equation \ref{eq:dnu_mag_B} is $\mathcal{I}$. Across all of the best fitting models, the variation in this parameter is well approximated by a Gaussian with a spread which is 30\% of the mean. If we take this as a conservative proxy for the model uncertainty on this parameter (given our stars share similar observable properties), we would expect a modeling error on the inferred field strength of order 30\%. We took this value forward as the model error, stating it in addition to the statistical error from the measurement of $\dnumag$, noting this is a very conservative estimate. 

\subsubsection{Mass}
The presence of core convection during the main sequence is dependent on stellar mass, requiring the star to have a mass greater than $\approx$1.1 M$_{\odot}$ \citep{1990sse..book.....K}. A magnetic field could then be driven in this convection zone and remain on the RGB in fossil form. Should significant magnetic perturbations only be measured in stars of mass > 1.1M$_{\odot}$, we may take this as evidence that core magnetic fields in red giants are the remains of those generated in a convection zone on the main sequence. However, as shown in \citet{li_magnetic_2022} and \cite{2023A&A...680A..26L}, stars with masses below 1.1M$_{\odot}$ can still develop small convective cores on the main sequence, due to the burning of $^3$He and $^{12}$C outside of equilibrium. There-in the authors find that the core sizes are not large enough to reach the hydrogen burning shell on the red giant branch. As such, modes would be less sensitive to the presence of the field and therefore require a larger field strength to produce a shift of similar scale. Assuming field strength does not depend on stellar mass, we should then find that measurable magnetic perturbations in stars with M < 1.1 M$_{\odot}$ are broadly smaller in magnitude.

\begin{figure}
    \centering
    \includegraphics[width=1\linewidth]{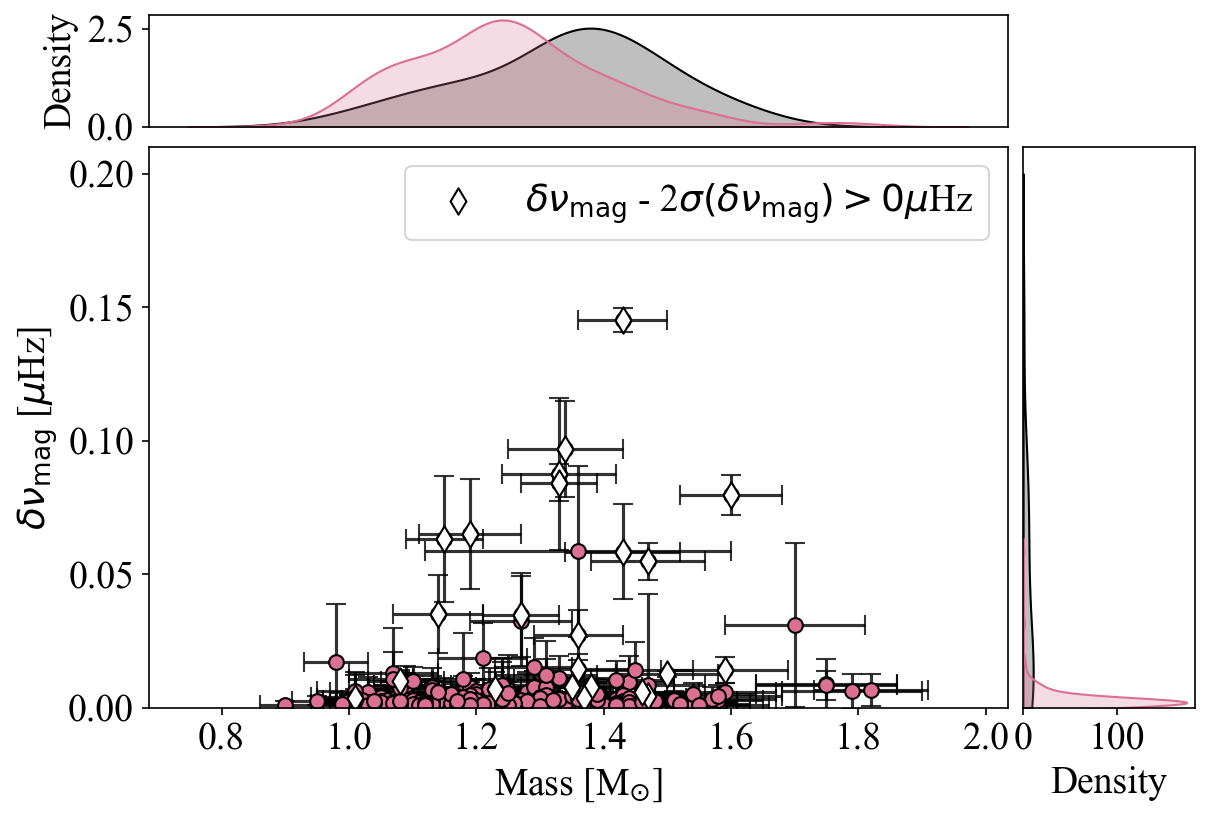}
    \caption{$\dnumag$ as a function of stellar mass. In the central panel the 23 stars with magnetic splitting values that are at least 2$\sigma$ from zero are shown in white diamonds, the remaining measurements are pink circles. The top panel shows KDEs of the mass distribution of the 23 stars in black and the remaining targets in pink. The right-most panel shows KDEs of the distribution of $\dnumag$, with the same colour-scheme.}
    \label{fig:dnumag mass}
\end{figure}

Figure \ref{fig:dnumag mass} shows $\dnumag$ as a function of stellar mass (mass values from Y18). We note that $\dnumag$ values in excess of 0.04$\mu$Hz only begin to appear at masses larger than $\approx$ 1.1 M$_{\odot}$. However, this is not the case for small but significant magnetic splittings at values below 0.04$\mu$Hz. In this regime stars span the full range in stellar mass present in our population. This could be consistent with a field driven in the small main sequence convection zone caused by the burning of $^3$He and $^{12}$C outside of equilibrium. Given that these targets have not previously been published due to the lack of asymmetry, this highlights a potential detection bias when manually selecting targets. This result could also be the signature of a systematic underestimate on stellar mass for the stars with significant $\dnumag$ at M < 1.1M$_{\odot}$. However, comparisons to masses derived from eclipsing binaries have shown masses from asteroseismic scaling relations are likely to be systematically overestimated rather than underestimated \citep{2016ApJ...832..121G, 2018MNRAS.476.3729B, 2018MNRAS.478.4669T, 2022ApJ...927..167L}.

\subsubsection{Magnetic Field Strengths}

To determine the value of $\Brms$ that would reproduce the magnetic splitting we measured, we utilized the best fitting models from the grid outlined in section \ref{sec:models}. From these models we calculated the value of $\mathcal{I}$ for each star. We then drew 1000 samples from the posterior distributions on $\dnumag$ and $\numax$ and calculated the mean field strength, $\Brms$, according to equation \ref{eq:dnu_mag_B}. The reported field strength for each star is taken as the mean of the resulting distribution on $\Brms$. Uncertainties are the standard deviation on this distribution plus the expected uncertainty from the models (see section \ref{sec:models}) added in quadrature. 

The distribution of $\Brms$ peaks at zero, reflecting the measured magnetic splitting. For the stars with a field value at least 2$\sigma$ from zero, 30\% of the population have $\Brms$ < 30kG. The distribution then tails with increasing field strength to the maximum at 169.4 $\pm$ 51 kG, occurring in KIC5696081.

\begin{figure}
    \centering
    \includegraphics[width=1\linewidth]{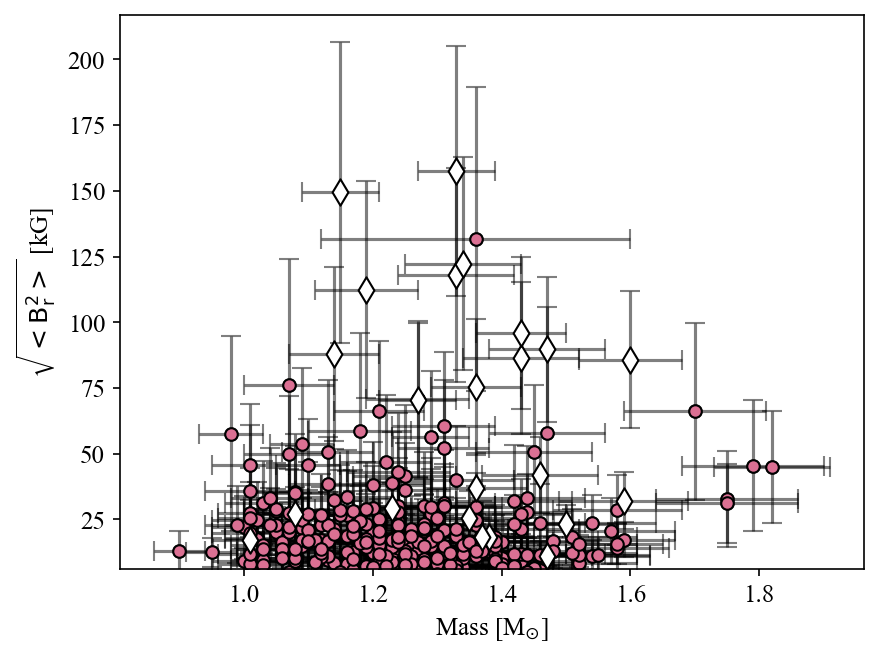}
    \caption{Best fitting $\sqrt{\Brms}$ as a function of stellar mass. The 23 stars with magnetic splitting values at least 2$\sigma$ from zero are marked with white diamonds. The remaining stars are marked with pink circles.}
    \label{fig:Brms mass}
\end{figure}

There is no significant correlation between the measured core field strengths and stellar mass (see figure \ref{fig:Brms mass}). At the low mass end of the distribution, the scale of the spread appears larger in $\Brms$ than $\dnumag$. This is a consequence of the $\numax^{-3}$ dependence on $\dnumag$. For a given field strength, a larger $\numax$ (preferentially occurring in lower mass stars) implies a smaller magnetic splitting. 

\begin{figure*}
    \centering
    \includegraphics[width=1\linewidth]{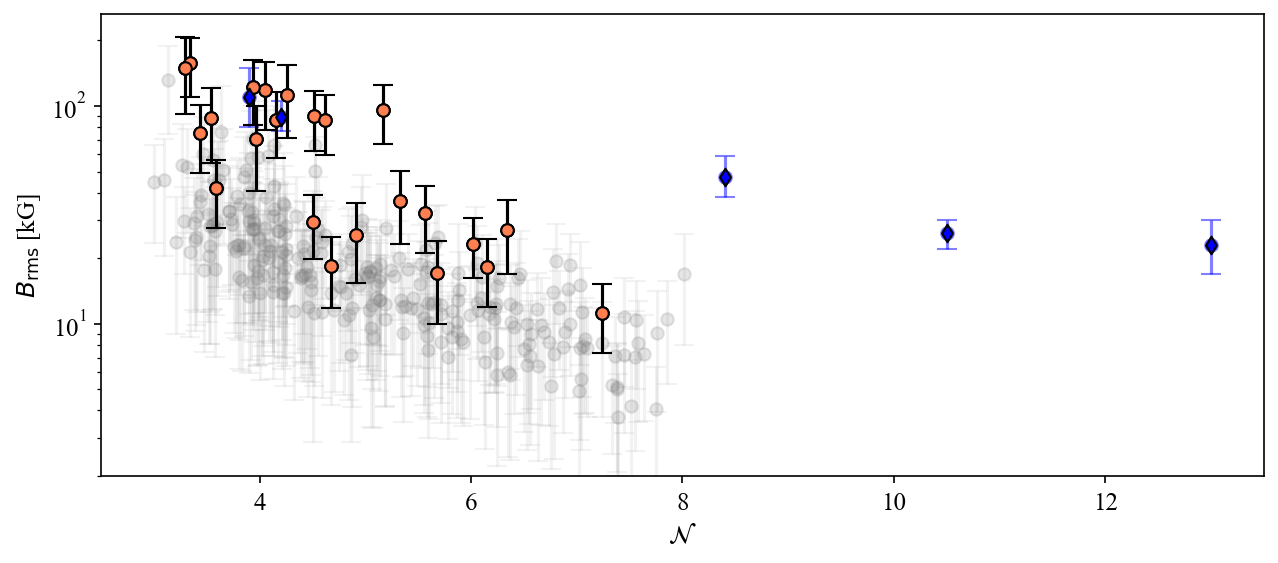}
    \caption{Best fitting $\sqrt{\Brms}$ as a function of mixed mode density, $\mathcal{N}$. Stars with $\dnumag$ - 2$\sigma(\dnumag)$ > 0 are shown in orange. The remaining measurements are in grey. Blue diamonds show values reported in L23 for stars that do not appear in this work. }
    \label{fig:Brms N}
\end{figure*}

L23 identified a decrease in the core field strength as stars evolve along the RGB, but caveat this with a note that there is significant scatter. We observe the same dependence in our larger set of stars, as is shown in figure \ref{fig:Brms N}. There-in the authors identified the decrease follows the decrease in the critical field strength, which sets an upper limit on the observable field strength. 

\section{Limitations}
Although the aim of this work was to catalogue a large number of stars such that we could start exploring population statistics, we were still subject to various detection biases. Firstly, we restricted ourselves to spectra where-in we could clearly identify all three components of the $\ell$ = 1 multiplets. Therefore, we are restricted to inclination in the approximate range 30$\degree$ $<$ $i$ < 60$\degree$. With the assumption that stellar inclination is isotropically distributed this limits us to $\approx$ 40\% of the possible sample of stars.

Secondly, targets were selected by manual identification of rotational splitting. This meant that we required spectra where-in the separation between mixed modes was significantly larger than the rotational splitting. This defined both the lower limit on $\numax$ (100 $\mu$Hz) and sets an upper limit on the $\dnurot$ for a given star. For a target with $\numax$ = 150 $\mu$Hz, $\Delta \Pi_{1}$ = 80 secs, the expected separation between adjacent mixed modes is approximately 3$\mu$Hz. Currently, the maximum recorded core rotation rate for a red giant is 0.95 $\mu$Hz \citep{gehan_core_2018}. Therefore, while our method is well suited to the range in core rotation previously reported in the literature, stars with rotation rates exceeding a few $\mu$Hz would not have appeared in our initial sample selection. 

Finally, We do not treat envelope rotation, which would introduce additional splitting in the p-dominated modes. Should the envelope rotation be significant, this could lead to an overestimate in our measurement of core rotational splitting. However, surface rotation rates in red giants are observed to be orders of magnitude smaller than the core rotation \citep{2013A&A...549A..75G}, such that they are unlikely to cause significant error. Non-standard stellar evolution (e.g. mergers) can cause rapid envelope rotation in red giants. However, best estimates for the prevalence of such non-standard rotators is on the order of 8\% \citep{2020A&A...639A..63G}. In our catalogue, $\approx$ 25\% of stars have core rotational splitting larger than 0.4$\mu$Hz. A study of the relation between envelope rotational splitting, core rotational splitting and magnetic parameters is reserved for future work. 

\section{Conclusions}

Exploiting the stretched period \'echelle, we have demonstrated how template matching can be used to construct initial estimates of the perturbations to dipole mode frequencies caused by core rotation and a magnetic field. We parameterise these using the magnetic splitting ($\dnumag$), a parameter dependent on field topology ($a$) and core rotational splitting ($\dnurot$). This allowed us to establish well-motivated priors for $\dnumag$ and $\dnurot$ in 334 low luminosity red giants.

Utilizing the information gained from the stretched \'echelles, we performed a full fit of the perturbed asymptotic expression to the power spectrum. This allowed us to jointly constrain $\Delta \Pi_{1}$, $q$, $\epsilon_{\mathrm{g}}$, $\delta \nu_{01}$, $\dnurot$, $a$ and $\dnumag$ in 302 targets. We found that not accounting for the magnetic perturbation when measuring the rotational splitting can lead to biased measurements when the magnetic perturbation is large ($\dnumag$ on the scale of 0.1$\mu$Hz). For the star with the largest value of $\dnumag$ the value of $\dnurot$ reported in G18 is 70\% smaller than the value we measured.

We identified a bimodality in the core rotation rates of the stars in our sample. The more populous peak is at $\dnurot$ = 0.32 $\mu$Hz, with the secondary at 0.47 $\mu$Hz. The location and size of this secondary peak appears to be mass dependent. We found the distribution also evolves with $\mathcal{N}$, with the upper limit on core rotation increasing with increasing $\mathcal{N}$. Assuming that in each $\mathcal{N}$ range the most rapidly rotating stars belong to a secondary population, the observed increase in core rotation rate would imply a relation of the form $\Omega_{\mathrm{core}}$ $\propto$ $R_{*}^{0.8}$. This is much closer to the predictions in \citet{2014ApJ...788...93C}, suggesting in these stars the evolution of core rotation could be reproduced using a combination of rotational and magnetic instabilities.

We measured a magnetic splitting that is at least 2$\sigma$ from zero in 8\% of the total sample (23 stars). Strong asymmetry was only present in 57\% of these targets (4\% of the full catalogue). For the stars with a clear detection of magnetic splitting, the topology parameter is not uniformly populated. A large percentage (35\%) have values of $a$ exceeding 0.5, identifiable with an architecture with the field more concentrated at the poles than the equator. Another large group (30\% of stars with significant magnetic splitting) have values of $a$ below -0.2, consistent with a field concentrated near the equator.

We did not observe any correlation between magnetic and rotational parameters, and so are unable to comment on whether the additional angular momentum transport is directly related to the magnetic fields we measured. 

Although the largest magnetic splittings we measured were in stars with masses greater than 1.1 M$_{\odot}$, magnetic splittings inconsistent with zero were measured in stars with masses from 1.03M$_{\odot}$ to 1.6M$_{\odot}$. This suggests that a main sequence convective core may not be the only channel for generating stable magnetic fields that are observed in fossil form on the red giant branch.

For the targets in which we measured significant magnetic splittings, the field strengths are on the order of tens of kG, with the number of detections decreasing with increasing field strength. The maximum value we measured was 169.4kG in KIC5696081. We confirm the tentative conclusion made in L23 that measurable field strengths decrease as stars evolve. 

\section*{Acknowledgements}
E.J.H., W.J.B. and G.R.D. acknowledge the support of Science and Technology Facilities Council. M.B.N. acknowledges support from the UK Space Agency. JMJO acknowledges support from NASA through the NASA Hubble Fellowship grant HST-HF2-51517.001, awarded by STScI, which is operated by the Association of Universities for Research in Astronomy, Incorporated, under NASA contract NAS5-26555. The authors acknowledge use of the Blue-BEAR HPC service at the University of Birmingham. This paper includes data collected by the Kepler mission and obtained from the MAST data archive at the Space Telescope Science Institute (STScI). Funding for the Kepler mission is provided by the NASA Science Mission Directorate. This work has made use of data from the European Space Agency (ESA) mission Gaia (\url{https://www.cosmos.esa.int/web/gaia}), processed by the Gaia Data Processing and Analysis Consortium (DPAC, \url{https://www.cosmos.esa.int/web/gaia/dpac/consortium}). Funding for the DPAC has been provided by national institutions, in particular the institutions participating in the Gaia Multilateral Agreement. This paper has received funding from the European Research Council (ERC) under the European Union’s Horizon 2020 research and innovation programme (CartographY GA. 804752). S.D. and J.B. acknowledge support from the Centre National d’Etudes Spatiales (CNES).

%%%%%%%%%%%%%%%%%%%%%%%%%%%%%%%%%%%%%%%%%%%%%%%%%%
\section*{Data Availability}
Table 1 is available at the CDS. Lightcurves used in this work are available at MAST. Other intermediate data products will be made available upon reasonable request. 

\begin{landscape}
\begin{table}
    \centering
    \begin{tabular}{c|c|c|c|c|c|c|c|c|c|c|c}
        \hline
        KIC ID & $\nu_{\mathrm{max}}$ ($\mu$Hz) & $\Delta \nu$ ($\mu$Hz) & q & $p_L$/10$^{-4}$ (Hz$^{2}$) & $p_D$/10$^{-4}$ & $\Delta \Pi_{1}$ (s) & $\epsilon_{\mathrm{g}}$ & $\delta \nu_{01}$ ($\mu$Hz) & $\dnu_\mathrm{mag}$ ($\mu$Hz) & $a$ & $\dnurot$ ($\mu$Hz) \\
        \hline
        10001728 & 138.8 $\pm$ 1.2 & 12.38 $\pm$ 0.01 & 0.1769 $\pm$ 0.0019 & 9.1 $\pm$ 0.9 & 75.2 $\pm$ 1.0 & 80.1 $\pm$ 0.02 & 0.931 $\pm$ 0.019 & 6.391 $\pm$ 0.012 & 0.0059 $\pm$ 0.0061 & 0.02 $\pm$ 0.29 & 0.3602 $\pm$ 0.0023\\
        10006097 & 142.9 $\pm$ 1.0 & 11.96 $\pm$ 0.01 & 0.1413 $\pm$ 0.0009 & 11.1 $\pm$ 0.9 & 63.4 $\pm$ 0.9 & 80.94 $\pm$ 0.01 & 0.827 $\pm$ 0.009 & 7.421 $\pm$ 0.009 & 0.0017 $\pm$ 0.0019 & 0.37 $\pm$ 0.32 & 0.3739 $\pm$ 0.0009\\
        10014959 & 176.7 $\pm$ 0.7 & 14.15 $\pm$ 0.0 & 0.1493 $\pm$ 0.0007 & 13.2 $\pm$ 1.2 & 72.2 $\pm$ 1.2 & 85.24 $\pm$ 0.01 & 0.864 $\pm$ 0.006 & 7.979 $\pm$ 0.01 & 0.004 $\pm$ 0.0031 & -0.23 $\pm$ 0.2 & 0.5358 $\pm$ 0.0008\\
        10063214 & 161.6 $\pm$ 0.8 & 12.85 $\pm$ 0.01 & 0.1381 $\pm$ 0.0007 & 12.0 $\pm$ 1.1 & 64.4 $\pm$ 1.1 & 81.13 $\pm$ 0.01 & 0.839 $\pm$ 0.006 & 7.342 $\pm$ 0.009 & 0.0025 $\pm$ 0.0024 & 0.02 $\pm$ 0.36 & 0.3637 $\pm$ 0.0005\\
        10068556 & 185.9 $\pm$ 0.4 & 13.97 $\pm$ 0.01 & 0.141 $\pm$ 0.001 & 11.1 $\pm$ 1.2 & 69.6 $\pm$ 1.2 & 81.48 $\pm$ 0.02 & 0.88 $\pm$ 0.015 & 7.032 $\pm$ 0.01 & 0.0095 $\pm$ 0.0097 & -0.08 $\pm$ 0.29 & 0.2316 $\pm$ 0.0008\\
        10078979 & 236.8 $\pm$ 1.3 & 18.05 $\pm$ 0.01 & 0.1825 $\pm$ 0.0007 & 16.3 $\pm$ 1.5 & 93.7 $\pm$ 1.4 & 90.73 $\pm$ 0.01 & 0.915 $\pm$ 0.004 & 10.071 $\pm$ 0.015 & 0.0019 $\pm$ 0.0028 & 0.11 $\pm$ 0.37 & 0.2009 $\pm$ 0.0011\\
        10149324 & 156.0 $\pm$ 0.7 & 12.48 $\pm$ 0.0 & 0.1344 $\pm$ 0.0009 & 12.3 $\pm$ 1.0 & 62.9 $\pm$ 1.0 & 83.3 $\pm$ 0.01 & 0.823 $\pm$ 0.007 & 7.53 $\pm$ 0.009 & 0.0078 $\pm$ 0.0029 & -0.41 $\pm$ 0.08 & 0.3282 $\pm$ 0.0005\\
        10198496 & 242.2 $\pm$ 1.0 & 17.76 $\pm$ 0.01 & 0.1544 $\pm$ 0.0006 & 13.5 $\pm$ 1.6 & 86.1 $\pm$ 1.5 & 89.27 $\pm$ 0.01 & 0.892 $\pm$ 0.004 & 9.681 $\pm$ 0.013 & 0.0018 $\pm$ 0.0033 & -0.03 $\pm$ 0.34 & 0.361 $\pm$ 0.001\\
        10199289 & 248.9 $\pm$ 1.1 & 18.49 $\pm$ 0.01 & 0.1817 $\pm$ 0.0006 & 13.9 $\pm$ 1.5 & 97.6 $\pm$ 1.4 & 91.4 $\pm$ 0.02 & 0.917 $\pm$ 0.007 & 9.995 $\pm$ 0.014 & 0.0087 $\pm$ 0.0068 & -0.19 $\pm$ 0.37 & 0.2693 $\pm$ 0.0009\\
        10274410 & 198.2 $\pm$ 0.8 & 14.88 $\pm$ 0.01 & 0.1601 $\pm$ 0.0009 & 14.1 $\pm$ 1.3 & 75.8 $\pm$ 1.3 & 83.84 $\pm$ 0.01 & 0.923 $\pm$ 0.004 & 7.202 $\pm$ 0.013 & 0.0012 $\pm$ 0.0014 & 0.35 $\pm$ 0.39 & 0.2657 $\pm$ 0.0009\\
        \hline
    \end{tabular}
    \caption{Example of the seismic parameters calculated according to the methodology detailed in section \ref{sec: methods}. The full catalogue is available online.}
    \label{tab:example_of_results}
\end{table}
\end{landscape}

%%%%%%%%%%%%%%%%%%%% REFERENCES %%%%%%%%%%%%%%%%%%

% The best way to enter references is to use BibTeX:

\bibliographystyle{mnras}
\bibliography{example} % if your bibtex file is called example.bib

% Alternatively you could enter them by hand, like this:
% This method is tedious and prone to error if you have lots of references
%\begin{thebibliography}{99}
%\bibitem[\protect\citeauthoryear{Author}{2012}]{Author2012}
%Author A.~N., 2013, Journal of Improbable Astronomy, 1, 1
%\bibitem[\protect\citeauthoryear{Others}{2013}]{Others2013}
%Others S., 2012, Journal of Interesting Stuff, 17, 198
%\end{thebibliography}

%%%%%%%%%%%%%%%%%%%%%%%%%%%%%%%%%%%%%%%%%%%%%%%%%%

%%%%%%%%%%%%%%%%% APPENDICES %%%%%%%%%%%%%%%%%%%%%

\appendix

\section{Parameters for Injection Recovery Test}

\begin{table}
    \begin{center}
        
        \begin{tabular}{lcccccr}%{lcccr}% four columns, alignment for each
                \hline
                $\Delta \Pi_{1}$ (s)& $\epsilon_{g}$ & q & $\dnu_{\mathrm{mag}}$ ($\mu$Hz) & $a$ & $\dnu_{\mathrm{rot}}$ ($\mu$Hz)\\
                \hline
                80.39 & 0.8 &  0.124 & 0.195 & 0.47 & 0.33 \\
                
                \hline
        \end{tabular}
        \caption{Values of the asymptotic parameters used to construct a mock spectrum for KIC8684542.}
        \label{tab:recovery}
        \end{center}
\end{table}
Table \ref{tab:recovery} contains the parameters used to generate the fake spectrum used in section \ref{sec:priors gmodes}.

\section{Comparisons to Literature Values}\label{appendix lit}

\begin{figure*}
    \centering
    \includegraphics[width=1\linewidth]{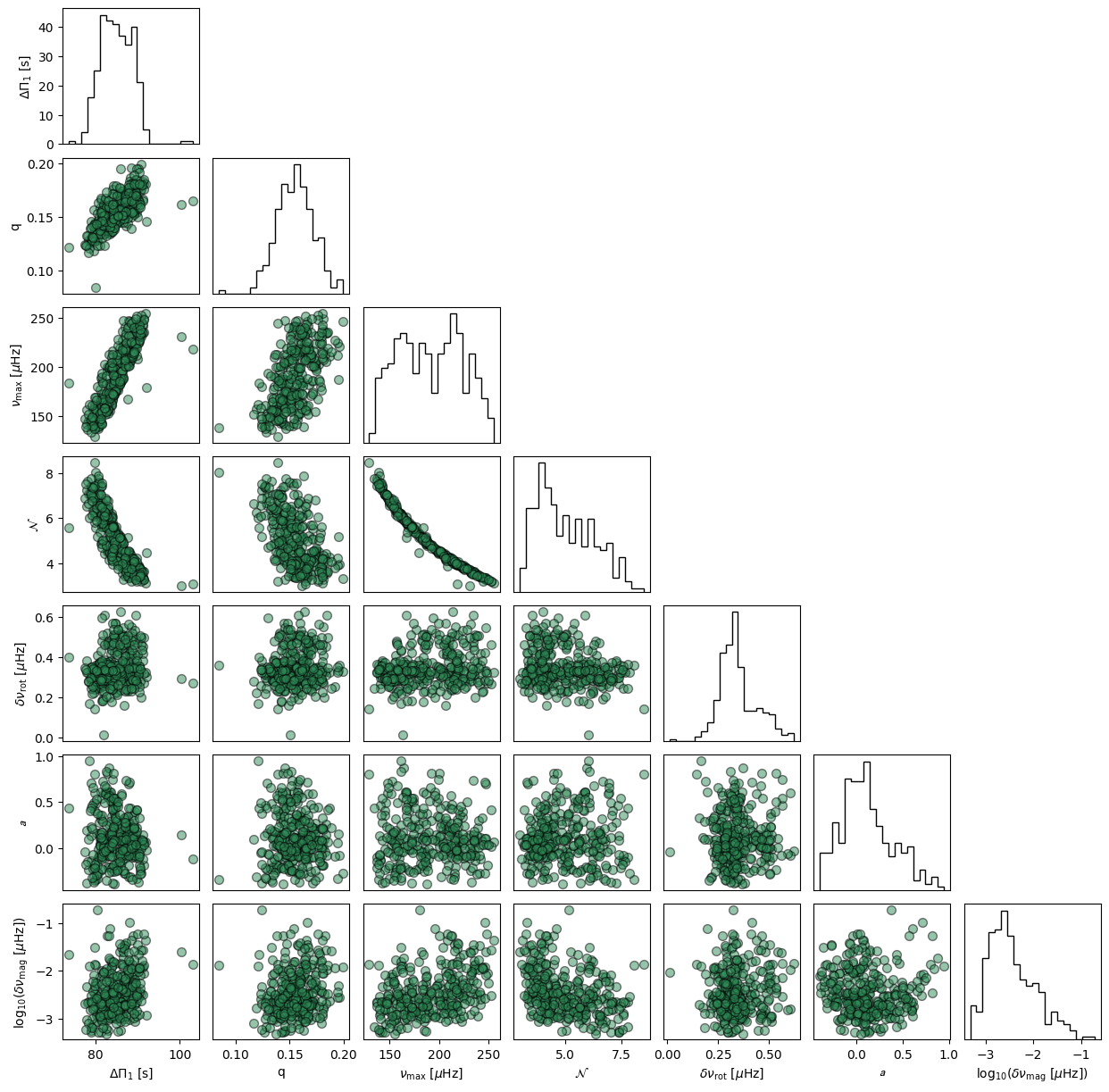}
    \caption{Corner plot showing the distribution of the asymptotic parameters, rotational splitting and magnetic parameters across all 302 stars. }
    \label{fig:obs corner}
\end{figure*}
\subsection{G-mode asymptotics}
\begin{figure}
    \centering
    \includegraphics[width=1\linewidth]{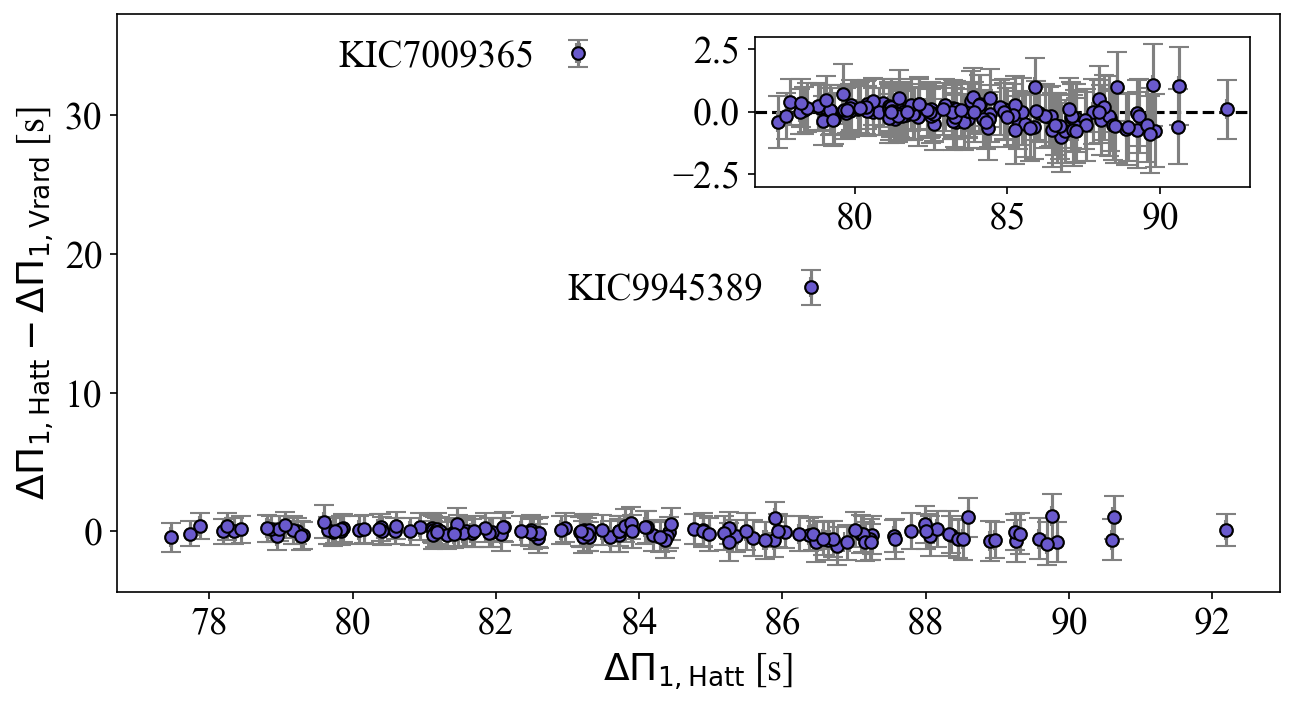}
    \caption{Difference in period spacing measured here and reported in V16.}
    \label{fig:diff period spacing}
\end{figure}

\begin{figure}
    \centering
    \includegraphics[width=1\linewidth]{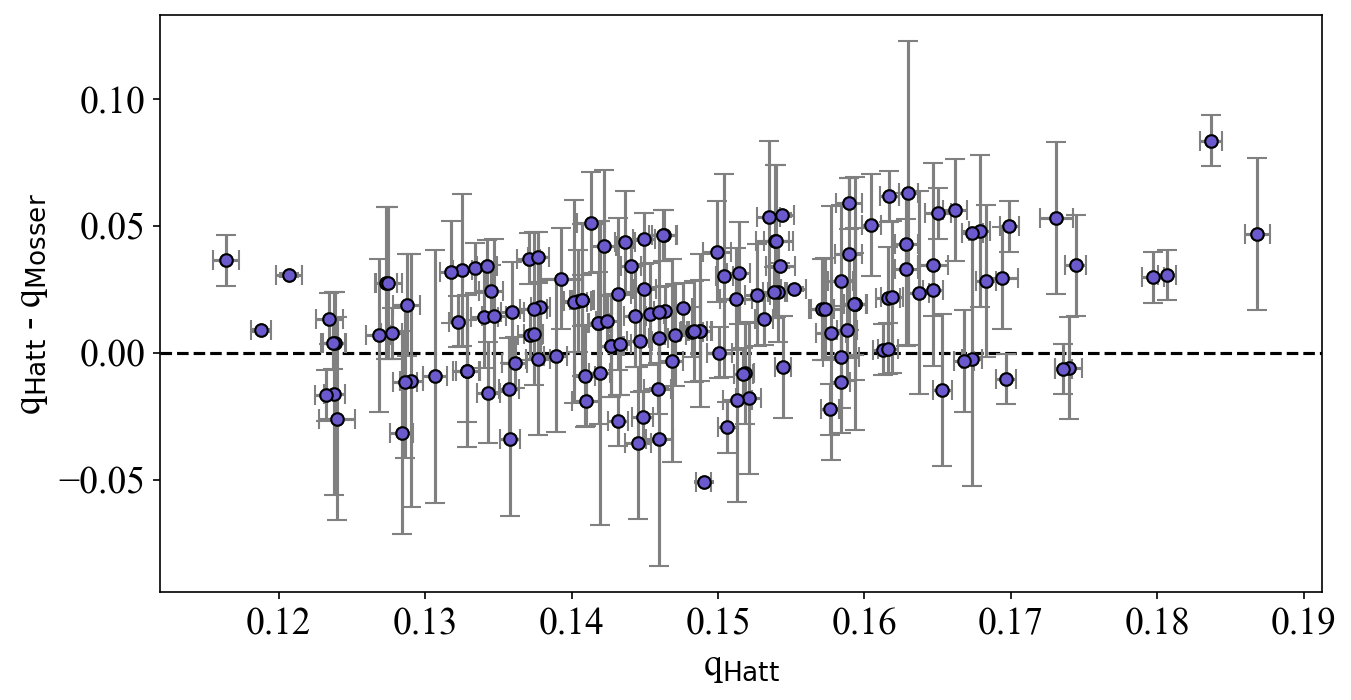}
    \caption{Difference in $q$ measured here and reported in M17.}
    \label{fig:diff coupling}
\end{figure}

We note here we manually inspected both the stretched echelles and the models of the power spectrum that result from our values of $\Delta \Pi_{1}$ and $q$ for each star in our sample, to confirm the values produced models in good agreement with the data. None-the-less, we include a comparison of our values with those from the literature below.
\cite{2016A&A...588A..87V} (V16) published measurements of g-mode asymptotics in 6100 red giants. \cite{mosser_period_2017} (M17) built on this catalogue to include the mode coupling parameter, $q$. In figures \ref{fig:diff period spacing} and \ref{fig:diff coupling} we show comparisons between our measurements of period spacing and coupling parameter (which we derive from equation \ref{eq: q}) and those from the aforementioned catalogues. The strong gridding in $q$ is a result of the methods used in M17. On average, the values of $q$ reported here are higher, with the mean offset being 10\%. This is below the average uncertainty reported in M17, which is 16\%. For a test star where-in the value of q in our work differs significantly from that in M17 we have included an example of the stretched \'echelle according to both results in the online materials. Our period spacing measurements are consistent with those reported in V16 in all but two cases, KIC 7009365 and KIC 9945389. The former case is one with a measurable magnetic signature, which was also reported in \citet{2023A&A...680A..26L}. For the latter star, the period spacing reported in V16 is 68.8s, which is substantially lower than the average in their sample (at 83.7s). The value reported here, however, is consistent with that average, at 86.4s. Though the mean of the posterior distribution when fitting for a magnetic shift in KIC 9945389 is not significantly different from zero, we note the posterior space is bimodal. A secondary set of solutions occurs at slightly higher period spacing and a significant magnetic shift. The stretched \'echelle for this star using the parameters published in this work and those in V16 can be found in the online materials. We note the uncertainties reported here are smaller than those in V16 by two orders of magnitude. They are, however, consistent with those reported in more recent studies exploiting similar fitting methods \citep{li_magnetic_2022,2023ApJ...954..152K,2023A&A...680A..26L}. 

\begin{figure}
    \centering
    \includegraphics[width=1\linewidth]{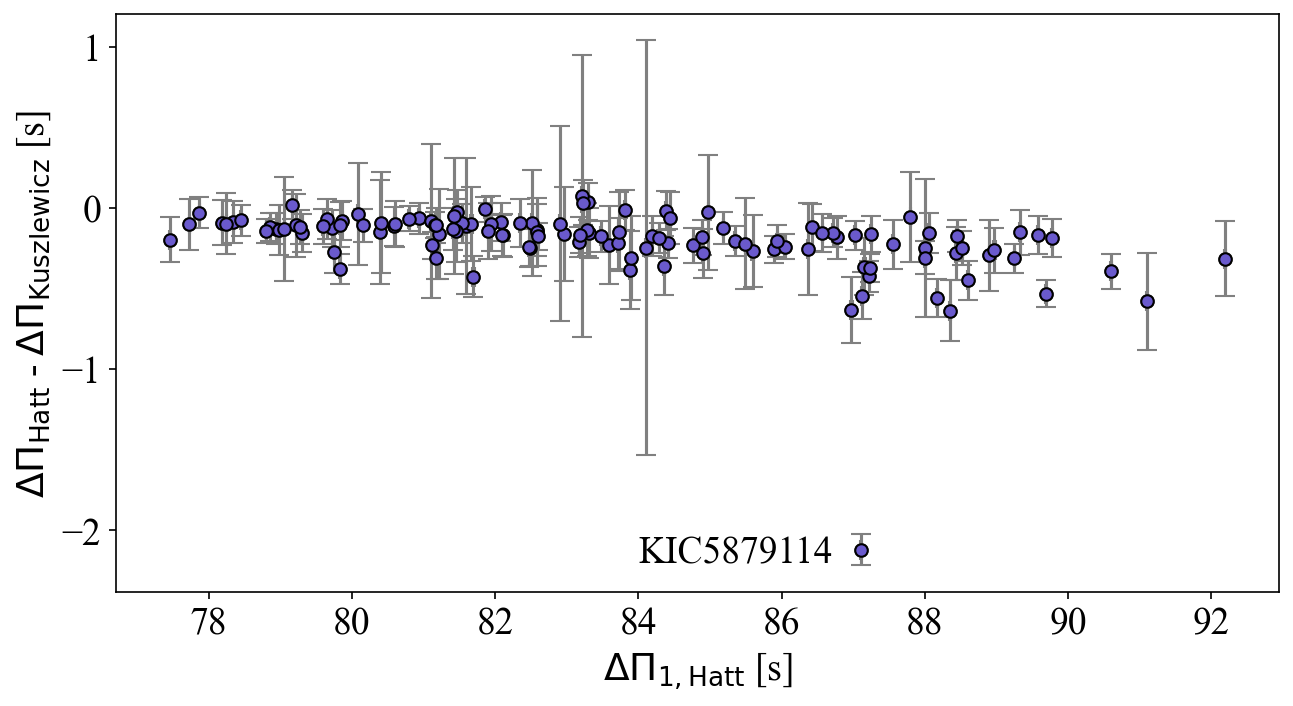}
    \caption{Difference in period spacing measured here and reported in K23.}
    \label{fig:diff DP K23}
\end{figure}

\begin{figure}
    \centering
    \includegraphics[width=1\linewidth]{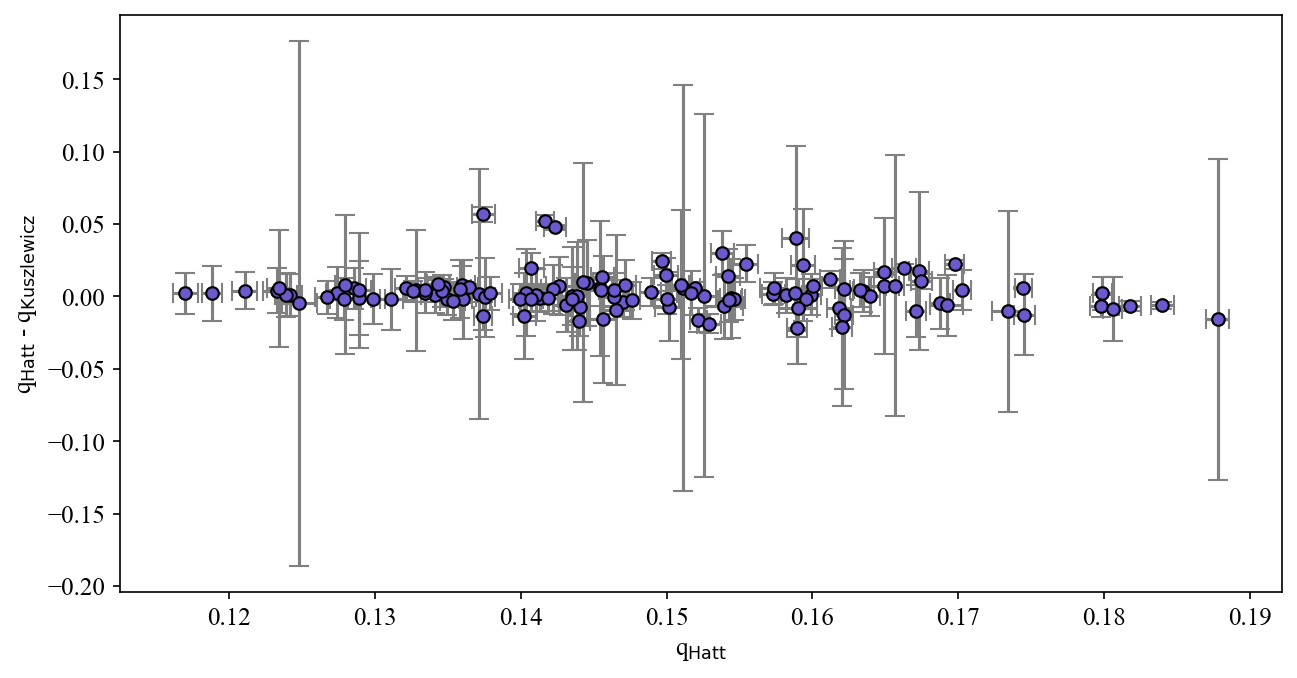}
    \caption{Difference in $q$ measured here and reported in K23.}
    \label{fig:diff q K23}
\end{figure}

Using a different method involving forward modelling mixed mode frequencies, \citet{2023ApJ...954..152K} (K23) reported the g-mode asymptotic parameters in 1074 \textit{Kepler} red giants. A comparison between our values and those in K23 can be seen in Fig. \ref{fig:diff DP K23} and \ref{fig:diff q K23}. There-in we find better agreement on both parameters than with V16 and M17. Noticeably, there is no trend or offset in q (the mean difference between the values reported here and those in K23 is 0.03\%).

\section{Example Corner Plots}
\begin{figure}
    \centering
    \includegraphics[width=1\linewidth]{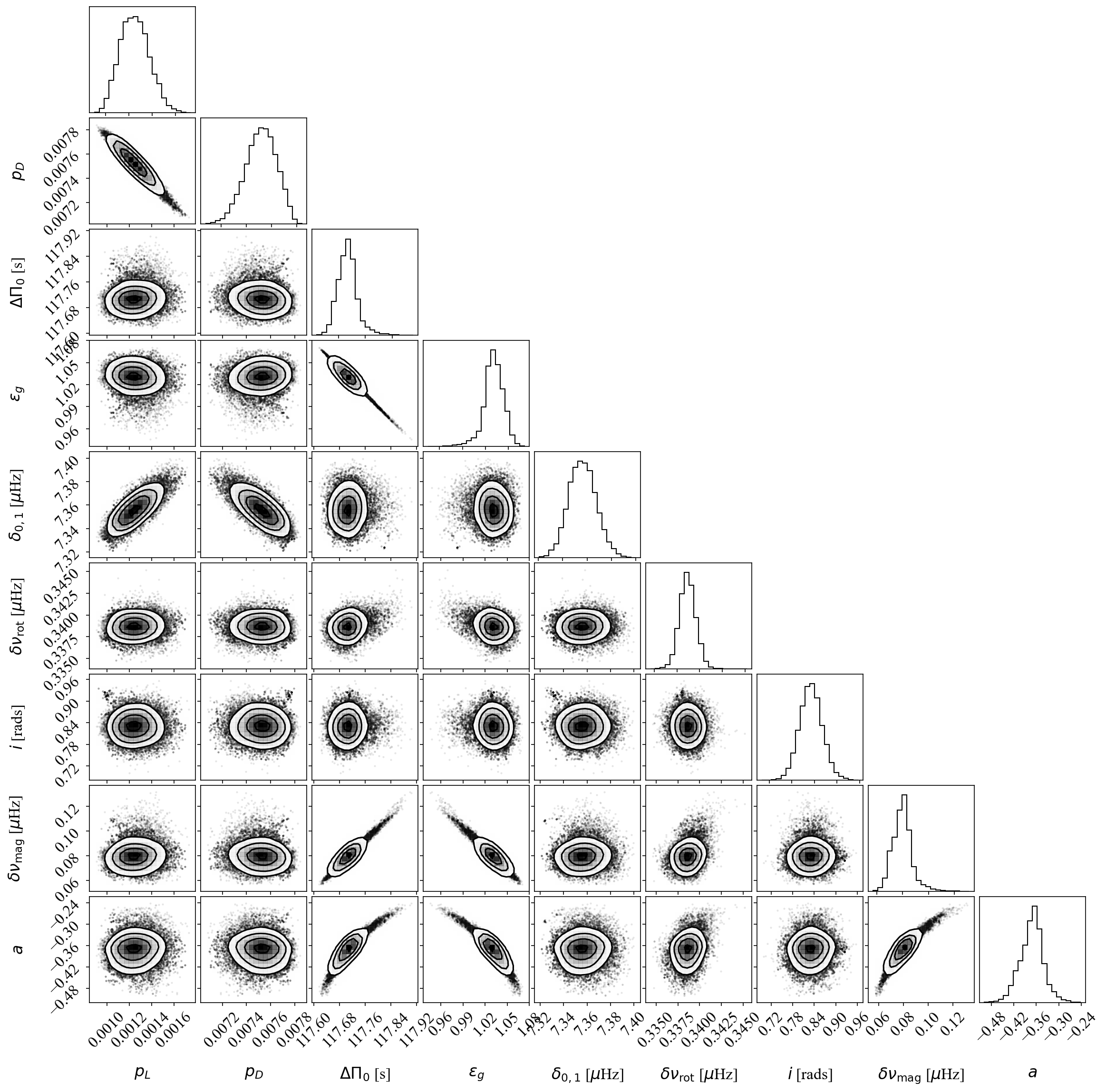}
    \caption{Corner plot of the asymptotic parameters of KIC11515377.}
    \label{fig:11515377 corner}
\end{figure}

\begin{figure}
    \centering
    \includegraphics[width=1\linewidth]{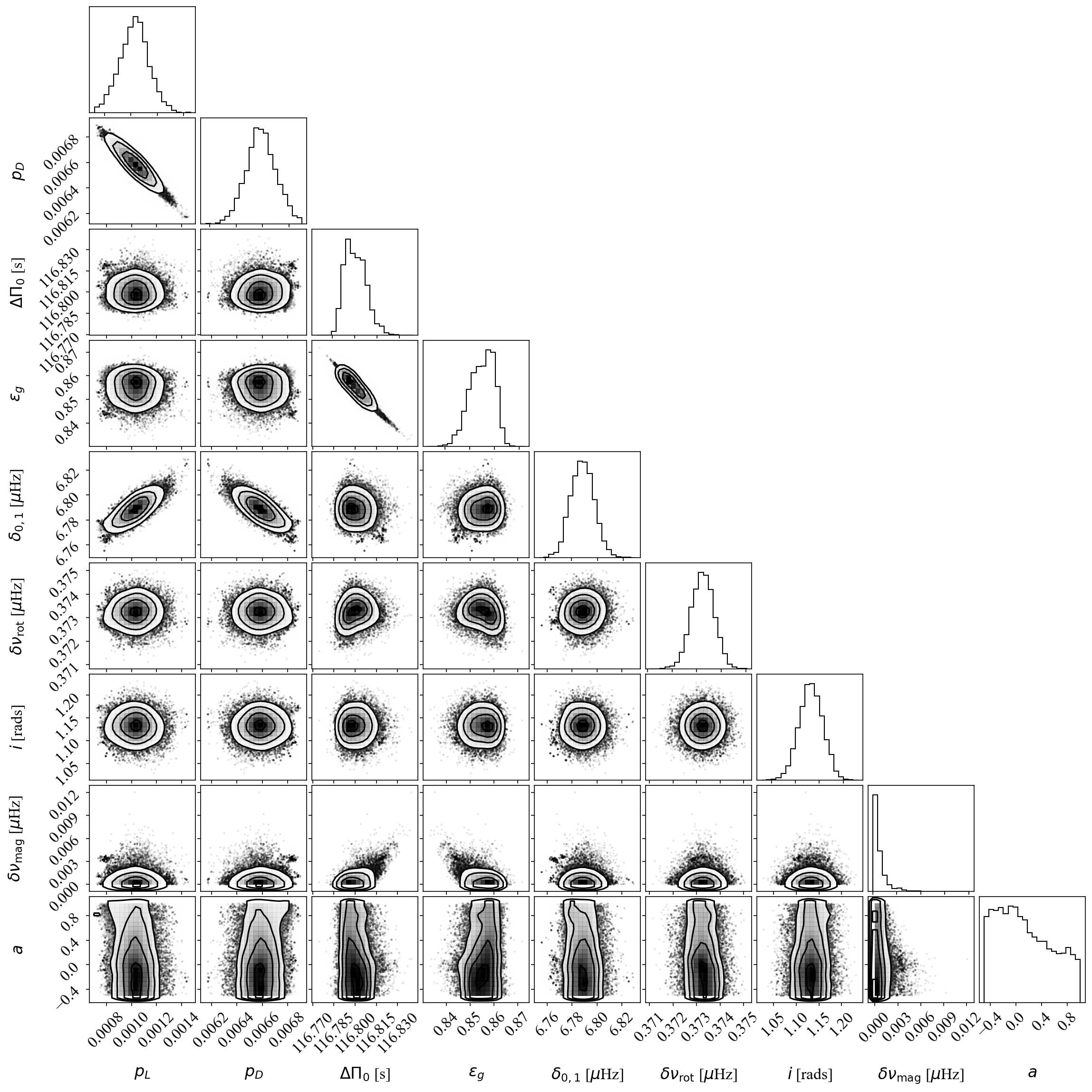}
    \caption{Corner plot of the asymptotic parameters of KIC7018212.}
    \label{fig:7018212 corner}
\end{figure}

Figures \ref{fig:11515377 corner} and \ref{fig:7018212 corner} show the corner plots of the parameters in the perturbed dipole mode model for two example stars. KIC 7018212 is a star in which no clear magnetic signature is measured, while KIC 11515377 has a clear signature of magnetic splitting.

\section{Establishing the Significance of Bimodality in Core Rotation}\label{appendix:bimodality}
\begin{figure}
    \centering
    \includegraphics[width=1\linewidth]{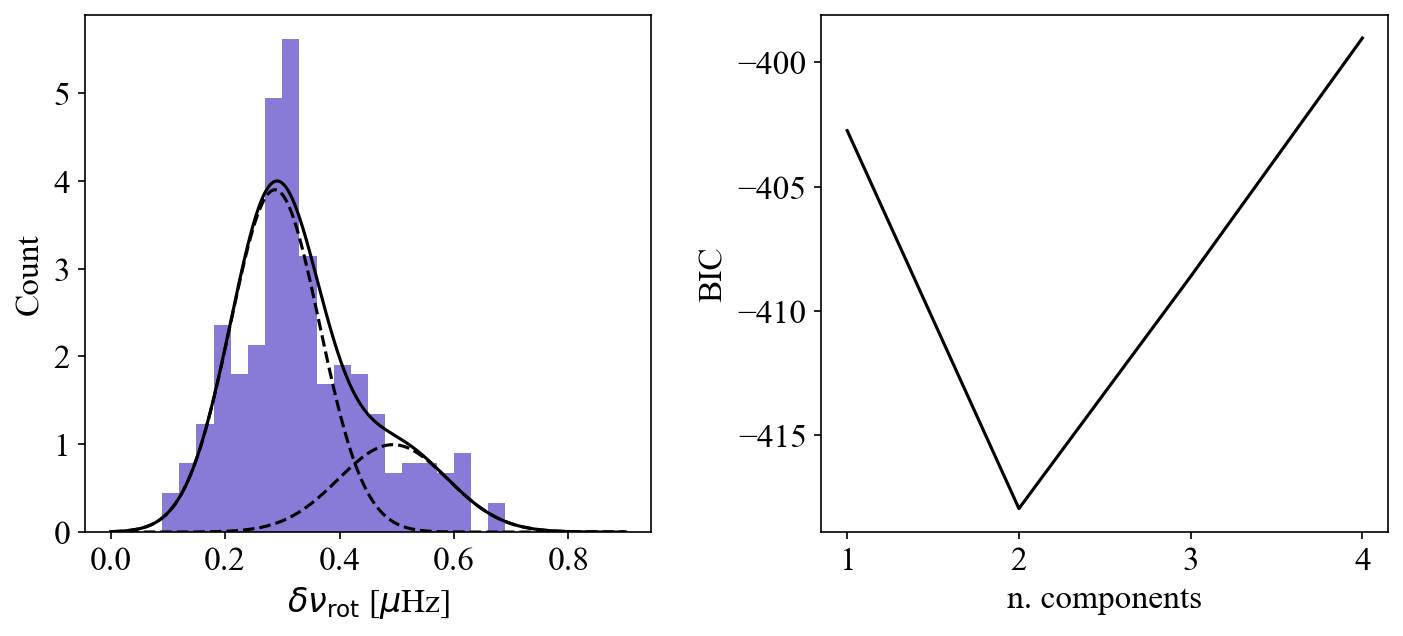}
    \caption{Left panel: Distribution of core rotational splitting for stars with $\nu_{\mathrm{max}}$ > 135$\mu$Hz, as reported in G18 in blue. Black lines represent a fit of two Gaussian distributions to the data. The individual components of this fit are represented by dotted lines. Right panel: The Bayesian information criterion (BIC) as a function of the number of Gaussian components used the fit the distribution. The minimum of this curve is for 2 components.}
    \label{fig:BIC 2018}
\end{figure}

\begin{figure}
    \centering
    \includegraphics[width=1\linewidth]{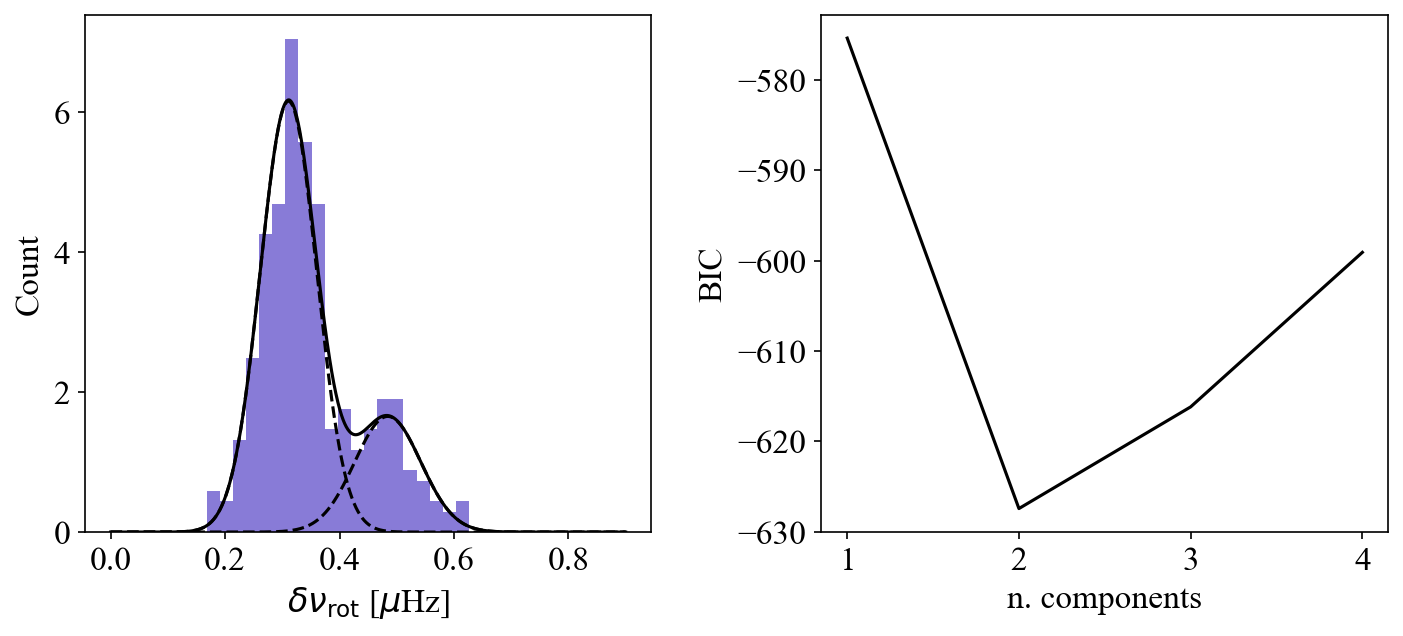}
    \caption{As in Fig. \ref{fig:BIC 2018}, but with the core rotational splitting measured in this work. Again the 2 component fit has the lowest BIC.}
    \label{fig:BIC 2023}
\end{figure}

To evaluate the significance of the apparent secondary peak in the distribution of $\dnurot$, we tested fitting different numbers of Gaussians to the distribution of core rotational splitting in G18 (for stars with $\nu_{\mathrm{max}}$ > 135$\mu$Hz, the minimum in our sample) and to the measurements reported here. To quantify which number of components best represented the data, while including a penalty for models with arbitrarily large numbers of free parameters (i.e. those that are over-fitting the data), we used the Bayesian information criterion \citep[BIC,][]{1978AnSta...6..461S}. This is defined as,

\begin{equation}
    \mathrm{BIC} = kln(n) - 2ln(\hat{L}),
\end{equation}
where $\hat{L}$ is the maximum of the likelihood for a given model, $n$ is the number of observations and $k$ is the number of parameters in the model. Models that minimize the BIC are thus preferred. We found in both the sub-sample from G18 and our measurements the value of the BIC was lower for the fit with two components, rather than one (see Fig. \ref{fig:BIC 2018} and \ref{fig:BIC 2023}). For the models with 2 Gaussian components, the means of the components were also consistent across the sub-sample from G18 ($\mu_{\mathrm{lower}}$ = 0.29$\mu$Hz, $\mu_{\mathrm{higher}}$ = 0.49$\mu$Hz) and the measurements reported here ($\mu_{\mathrm{lower}}$ = 0.31$\mu$Hz, $\mu_{\mathrm{higher}}$ = 0.48$\mu$Hz).

%%%%%%%%%%%%%%%%%%%%%%%%%%%%%%%%%%%%%%%%%%%%%%%%%%

% Don't change these lines
\bsp	% typesetting comment
\label{lastpage}
\end{document}